\documentclass[12pt]{amsart}
\usepackage{amssymb}
\usepackage{amsmath}
\usepackage{amsfonts}
\usepackage{mathrsfs}
\usepackage{graphicx}
\usepackage{placeins}
\usepackage{color}
\usepackage[onehalfspacing]{setspace}
\usepackage{caption}
\usepackage{subcaption}
\usepackage{natbib}
\usepackage{enumerate}
\usepackage[utf8]{inputenc}
\usepackage[charter,cal=cmcal]{mathdesign}
\usepackage{epigraph}
\usepackage[colorlinks=true,citecolor=blue,urlcolor=blue,pdfpagemode=UseNone,pdfstartview=FitH]{hyperref}
\usepackage{apptools}
\AtAppendix{\counterwithin{lemma}{section}}

\makeatletter
\def\section{\@startsection{section}{1}
	\z@{1.0\linespacing\@plus\linespacing}{.8\linespacing}{\Large}}

\def\subsection{\@startsection{subsection}{2}
	\z@{.8\linespacing\@plus.7\linespacing}{.7\linespacing}{\large}}

\def\subsubsection{\@startsection{subsubsection}{3}
	\z@{.5\linespacing\@plus.7\linespacing}{-.5em}{\normalfont\bfseries}}
\makeatother

\numberwithin{equation}{section}

\newtheorem{theorem}{Theorem}[section]
\newtheorem{lemma}{Lemma}[section]

\newtheorem{corollary}{Corollary}[section]

\theoremstyle{definition}
\newtheorem{definition}{Definition}[section]

\theoremstyle{definition}

\theoremstyle{definition}

\DeclareTextFontCommand{\bi}{%
	\fontseries\bfdefault 
	\itshape
}

\DeclareMathOperator{\Var}{Var}
\DeclareMathOperator{\Cov}{Cov}
\DeclareMathOperator{\diag}{diag}

\title{}
\begin{document}
	\vspace*{5ex minus 1ex}
	\begin{center}
		\Large \textsc{Some Impossibility Results for Inference With Cluster Dependence with Large Clusters}
		\bigskip
	\end{center}

	\date{%
		\today%
	}

	\vspace*{3ex minus 1ex}
	\begin{center}
		\medskip

		Denis Kojevnikov and Kyungchul Song\\
		\textit{Tilburg University and University of British Columbia}\\
		\medskip
	\end{center}

	\fontsize{13}{14} \selectfont

	\begin{abstract}
		{\footnotesize
			This paper focuses on a setting with observations having a cluster dependence structure and presents two main impossibility results. First, we show that when there is only one large cluster, i.e., the researcher does not have any knowledge on the dependence structure of the observations, it is not possible to consistently discriminate the mean. When within-cluster observations satisfy the uniform central limit theorem, we also show that a sufficient condition for consistent $\sqrt{n}$-discrimination of the mean is that we have at least two large clusters. This result shows some limitations for inference when we lack information on the dependence structure of observations. Our second result provides a necessary and sufficient condition for the cluster structure that the long run variance is consistently estimable. Our result implies that when there is at least one large cluster, the long run variance is not consistently estimable.}\bigskip

		{\footnotesize \noindent \textsc{Key words.} Consistent Discrimination; Local Dependence; Unknown Dependence Structure; Consistent Estimation of Long-Run Variance; Cluster Dependence; Log Likelihood Process \bigskip\ }

	\end{abstract}

	\thanks{We thank Michael Leung and James MacKinnon for valuable comments. We thank the Associate Editor and three anonymous referees for their constructive comments which helped us improve our paper's exposition and results. All errors are ours. Song acknowledges financial support from Social Sciences and Humanities Research Council of Canada. Corresponding Address: Kyungchul Song, Vancouver School of Economics, University of British Columbia, 6000 Iona Drive, Vancouver, V6T 1L4, Canada.}
	\maketitle

	\normalsize

	\section{Introduction}

	Statistical inference from data usually begins by imposing a form of a dependence structure on the data, by specifying which groups of observations exhibit strong within-group dependence. Various tools of asymptotic inference such as the law of large numbers and the central limit theorem are available for many typically imposed dependence structures. A standard case is the independence assumption or an assumption on time series dependence. However, it is well known that in the case of cross-sectional dependence, a researcher is often less confident about the correctness of the dependence structure used, despite its crucial role for inference.

	A popular way to deal with this challenge is to use cluster dependence modeling, where the dependence structure among observations within each cluster is left unspecified, while independence is imposed between observations from different clusters. The inference procedures when there are many clusters are well known and can be analyzed using standard methods of asymptotic inference. However, less is known about the case where there are large clusters, and the dependence structure within such a cluster is unknown. \cite{Cameron/Gelbach/Miller:08:ReStat} proposed a wild bootstrap procedure and showed by simulations that their tests perform well even when there are a small number of clusters. The robustness of this result was confirmed by \cite{MacKinnon/Webb:17:JAE} even when the sizes of the clusters are highly heterogeneous. This large cluster issue has also drawn interest in the literature of difference-in-differences when there are only few treated clusters (see \cite{Conley/Taber:11:ReStat}, \cite{Hagemann:19:JOE}, and \cite{MacKinnon/Webb:20:JOE}, and references therein). \cite{Djogbenou/MacKinnon/Nielsen:19:JOE} studied inference on regression models with clustered errors. They provided conditions for the cluster sizes so that asymptotic and bootstrap inferences are asymptotically valid. They showed that their conditions exclude the presence of a large cluster.

	There are several methods proposed to deal with the problem of inference with large clusters. \cite{Donald/Lang:07:ReStat} and \cite{Bester/Conley/Hansen:11:JOE} considered linear models and proposed inference where the asymptotic distribution of the long run variance estimator is fully known. This approach is related to the HAR (Heteroskedasticity-Autocorrelation Robust) inference of \cite{Kiefer/Vogelsang:2002:Eca} and \cite{Sun:14:Eca} in time series, which uses a normalization by an inconsistent long run variance estimator that has a stochastic limit.

	\cite{Ibragimov/Muller:10:JBES} proposed a $t$-test approach based on within-cluster estimators together with a $t$-distribution, where the degree of freedom in the $t$ distribution is determined by the number of clusters. They used the result of \cite{Bakirov/Szekely:05:ZNS} and showed that their approach is asymptotically valid, even if the variances of the cluster specific estimators are different across the clusters. \cite{Ibragimov/Muller:16:ReStat} extended these results to the problem of two-sample comparison and developed a testing procedure for the level of clustering.

	Some studies adopted the approach of randomized testing to deal with cluster dependence with large clusters. \cite{Canay/Romano/Shaikh:17:Eca} developed asymptotic inference procedures when the inference involves statistics whose limiting distribution satisfy symmetry properties. \cite{Hagemann:19:JOE} proposed randomized tests for treatment effects when there are only a small number of clusters. Like \cite{Ibragimov/Muller:10:JBES}, both proposals assumed large sample properties for within-cluster statistics. A recent work by \cite{Canay/Santos/Shaikh:21:ReStat} use the analogue between wild bootstrap and randomized tests, and provided conditions under which the wild bootstrap for cluster-dependent regression models is asymptotically valid when there are only a small number of clusters.

	Our paper focuses on observations with a cluster dependence structure and explores implications on statistical inference when there are large clusters. First, we show that when the sample consists of large clusters, the mean cannot be consistently discriminated if there is only one cluster, i.e., the researcher does not have any knowledge on the dependence structure of the data. Furthermore, when the observations form large clusters and within-cluster observations satisfy the uniform central limit theorem, a sufficient condition for the mean to be consistently discriminated at the rate of $\sqrt{n}$ is that the sample consists of at least two large clusters.

	This impossibility result has a significant implication in a setting where the researcher does not know the dependence structure of observations. In such a case, consistent discrimination of the mean is not possible with uniform-in-$P$ asymptotic size control. Note that \cite{Song:16:arXiv} proposed a randomized subsampling approach, and \cite{Leung:21:JAE} provided a set of general conditions for the approach to produce asymptotically valid inference. Both focus on a setting where no knowledge on the dependence structure is required. Among other things, their results show that the mean is consistently discriminated. Our impossibility result on consistent discrimination considers a setting where there is no uniform upper bound of the long run variance in the null model, and this setting is excluded by part of their conditions. Hence, their results do not contradict our impossibility result. 

	Our second result is concerned with consistent estimation of long run variances. More specifically, suppose that $X_n = [X_{n,1},...,X_{n,n}]^\top$ is a given random vector of dimension $n$, where each observation $X_{n,i}$ has the same mean $\mu$. Let us define the long-run variance of $X_n$ as follows:\footnote{Note that when there is a common shock, say, $C_n$, such as cluster-specific fixed effects with few clusters, the analysis in this paper carries over to this case with $\sigma_{LR}^2$ replaced by the conditional variance given common shock $C_n$. Our impossibility results do not depend on whether there is a common shock of this form in the data or not. For simplicity, we consider a setting without such cluster-specific fixed effects.}
	\begin{align}
		\label{LR var}
		\sigma_{LR}^2 = \Var\left( \frac{1}{\sqrt{n}}\sum_{i=1}^n X_{n,i} \right).
	\end{align}
    Recently, \cite{Hansen/Lee:19:JOE} derived an asymptotic distribution theory for clustered data, including a law of large numbers and a central limit theorem. One of their results presents a condition for the cluster structure that is necessary and sufficient for the weak law of large numbers to hold for the sample average of the clustered observations. Our paper shows that the same condition is in fact necessary and sufficient for the consistent estimability of the long run variance of the clustered observations as well. Our condition for the cluster structure also implies that when there is at least one large cluster, i.e., the researcher does not know the dependence structure on a nonnegligible portion of the data, the long-run variance is not consistently estimable. It is not hard to show that the existing cluster-robust variance estimators are inconsistent when the cluster structure is severely misspecified. However, to the best of our knowledge, it has not been known whether there exists any consistent estimator of the long run variance when there is a lack of knowledge on the dependence structure on a nonnegligible portion of the data. Our result gives a negative answer to this question.

	There has long been a strand of literature that studies impossibility of estimation and inference. (See, e.g., \cite{Bahadur/Savage:56:AMS}, \cite{Dufour:97:Eca}, \cite{Potscher:02:Eca}, \cite{Bertanha/Moreira:20:JOE}.) The impossibility of consistent estimation of a long run variance in this paper is related to \cite{Potscher:02:Eca} who established  a minimax risk lower bound for a general estimation problem. Among others, his result can be used to prove the impossibility of consistent estimation uniform in $P$ as shown in Corollary 3.2 there. However, we cannot apply this corollary in our setting, because our probability model is not indexed by a set of parameters fixed independently of the sample size, such as $\mathscr{H}$ in his paper. This stems from our setting where we have to deal with the \textit{joint distribution of the entire sample} whose dependence structure varies in the model as $n$ changes. \cite{Bertanha/Moreira:20:JOE} studied impossibility results of two types: indistinguishability of the null hypothesis from the alternative hypothesis and unbounded confidence sets. Their study of impossibility of the first type is related to impossibility of consistent discrimination of the mean in our paper. For this result, they assume that for each probability in the alternative hypothesis, there is a sequence of probabilities under the null hypothesis that weakly converge to this probability. Our setting does not satisfy this assumption in general. Hence, our result does not fall into their framework. \cite{Menzel:21:Eca} recently developed and verified the validity of a bootstrap procedure in multi-way clustered observations with two or more dimensions. Part of his results shows that it is not possible to consistently estimate the distribution of the cluster dependent observations. Our results are not the special case of his results, because our impossibility result holds for models that exclude the counterexample that he used to prove the impossibility result. In particular, our cluster dependence accommodates within-cluster heterogeneity in terms of marginal distributions and dependence structures.

	The rest of the paper is organized as follows. The next section studies the consistent discrimination of the mean. Section 3 is devoted to presenting the result of the impossibility of consistent estimation of the long run variance. In Section 4, we illustrate the implication of our results for the case of difference-in-difference models. In Section 5, we conclude. The mathematical proofs are found in the appendix.

	\section{Cluster Dependence}

	Let $X_n = [X_{n,1},...,X_{n,n}]^\top$ be a random vector with a joint distribution $P_n$ which belongs to the class of distributions $\mathcal{P}_n$. Throughout the paper, we assume that for each $P_n \in \mathcal{P}_n$,
	\begin{align*}
		\mathbf{E}[X_{n,1}] = \mathbf{E}[X_{n,2}] = ... = \mathbf{E}[X_{n,n}],
	\end{align*}
	and $\sigma_{LR}^2 < \infty$, where $\sigma_{LR}^2$ is defined in (\ref{LR var}). In many situations, the dependence structure is partially observed. Here we consider cluster dependence, where the dependence structure is entirely unknown within each cluster, and observations are independent between clusters. Let $N_{n,m}, m = 1,...,M_n$, be a partition of $N_n = \{1,...,n\}$ such that $|N_{n,m}| = n_m$ for each $m=1,...,M_n$, so that $\sum_{m=1}^{M_n} n_m = n$. Define $\mathcal{M}_n = \{N_{n,m}: m = 1,...,M_n\}$ and call it a \bi{cluster structure}. Throughout the paper we assume that $(X_{n,i})_{i \in N_{n,m}}$ are independent across $m$'s under all $P_n \in \mathcal{P}_n$, i.e., the joint distribution of $X_n$ has a cluster dependence structure. For future references, we define 
    \begin{align*}
        \overline X_{m,n} = \frac{1}{n_m}\sum_{i \in N_m} X_{n,i} \quad \text{ and } \quad \sigma_{n,m}^2 = \Var\left( \sqrt{n_m} \overline X_{m,n} \right),
    \end{align*}
    so that $\overline X_{m,n}$ represents the within-cluster mean of $X_{n,i}$'s and $\sigma_{n,m}^2$ represents the within-cluster long-run variance of $X_{n,i}$'s. 

	Our impossibility results rely on the assumption that the probability model, $\mathcal{P}_n$, includes Gaussian experiments with what we call local-to-independence common shocks. For each cluster $m=1,...,M_n$, and for $\delta > 0$ and $\sigma^2>0$, we define
    \begin{eqnarray*}
        \Sigma_{n,m}(\sigma^2,\delta) = \sigma^2\left(\left(1- \frac{\delta}{n_m}\right)I_{n_m} +  \frac{\delta}{n_m} \mathbf{1}_{n_m} \mathbf{1}_{n_m}^\top\right),
    \end{eqnarray*}
    where $I_{n_m}$ denotes the $n_m$-dimensional identity matrix and $\mathbf{1}_{n_m}$ is the $n_m$-dimensional column vector of ones. Let $\Sigma_n(\sigma^2,\delta)$ be the $n \times n$ block diagonal matrix whose $m$-th block is given by $\Sigma_{n,m}(\sigma^2,\delta)$. Suppose that $\Sigma_n(\sigma^2,\delta)$ is positive definite. Then, for each $\mu_n \in \mathbf{R}^n$, we denote $\Phi(\mu_n,\Sigma_n(\sigma^2,\delta))$ the multivariate normal distribution with mean $\mu_n$ and covariance matrix $\Sigma_n(\sigma^2,\delta)$. Define
	\begin{eqnarray*}
		\mathcal{P}_{n,\mathcal{N}} = \left\{\Phi(\overline \mu \mathbf{1}_n,\Sigma_{n}(\sigma^2, \delta)): \sigma^2>0, \delta \in (0,1] \text{ and }\overline \mu \in \mathbf{R} \right\}.
	\end{eqnarray*}
    The set $\mathcal{P}_{n,\mathcal{N}}$ represents a set of Gaussian models, where each member is a multivariate normal distribution with a common mean and an equal covariance. We call each $\Phi(\mu_n,\Sigma_n(\sigma^2,\delta))$ the \bi{local-to-independence common shock (LTIC) Gaussian distribution} with parameters $\sigma^2$ and $\delta$. This Gaussian distribution represents the cross-sectional dependence structure of $X_{n,i}$'s generated as follows:
    \begin{eqnarray*}
        X_{n,i} = \mu_{n,i} + \varepsilon_i \sqrt{1 - \frac{\delta}{n_m}} + \eta_m \sqrt{\frac{\delta}{n_m}}, \text{ whenever } i \in N_m,
    \end{eqnarray*}
    where $\mu_{n,i}$ is the $i$-th entry of $\mu_n$, $\varepsilon_i$'s are i.i.d.\ normal random variables with mean zero and variance $\sigma^2$, and $\eta_m$, $m=1,...,M_n$, are i.i.d.\ normal random variables with mean zero and variance $\sigma^2$, independent of $\varepsilon_i$'s. Each random variable $\eta_m$ represents a within-cluster ``common shock'', and creates the within-cluster global dependence among $X_{n,i}$'s. The influence of this common shock on the random variable $X_{n,i}$ diminishes at the rate of $\sqrt{n_m}$.
    
	\section{Consistent Discrimination of the Mean}

	\subsection{Consistent Discrimination of the Mean}

	Let us explore the consistent discrimination of the mean under the general cluster dependence structure.  We introduce the notion of consistent discrimination formally. Let $\mathcal{P}_n$ be a set of the distributions of $X_n \in \mathbf{R}^n$ such that $\mathbf{E}[X_{n,i}]$ is identical across $i$ for each $n \ge 1$. Let $\mathcal{P}_{n,0} = \{P_n \in \mathcal{P}_n: \mathbf{E}[X_{n,i}] = 0\}$, i.e., the set of probabilities under the null hypothesis of $\mathbf{E}[X_{n,i}] = 0$.  

	\begin{definition}
	    The mean of $X_{n,i}$, $n \ge 1$, is \bi{consistently discriminated at level $\alpha \in (0,1)$ in model $\mathcal{P}_n$}, if there is a sequence of (potentially randomized) tests $\{\varphi_n\}_{n \ge 1}$ such that 
		\begin{align*}
			\limsup_{n \rightarrow \infty} \mathbf{E}[\varphi_n(X_n)] \le \alpha, 
		\end{align*}
		along any sequence $P_{n,0} \in \mathcal{P}_{n,0}$, and 
		\begin{align*}
			\liminf_{n \rightarrow \infty}  \mathbf{E}[\varphi_n(X_n)] = 1,
		\end{align*}
		along any sequence $P_n \in \mathcal{P}_n$ such that $\liminf_{n \ge 1}\mathbf{E}[X_{n,i}] /\sigma_{LR} >0$.
	\end{definition}

	The following theorem shows that when the sample consists of nonnegligible clusters, a necessary condition for the consistent discrimination of the mean is that there exist at least two clusters.
	\begin{theorem}
		\label{thm: consistent discrimination}
		Suppose that $\mathcal{P}_{n,\mathcal{N}} \subset \mathcal{P}_n$ for each $n \ge 1$. Suppose further that $\alpha \in (0,1/2)$, and $M_n = 1$ for each $n \ge 1$. Then, the mean cannot be consistently discriminated at level $\alpha$.
	\end{theorem}

   The theorem implies that when we do not know the local dependence structure of the random variables (i.e., $M_n = 1$), it is not possible to consistently discriminate the mean.

    \subsection{Consistent $\sqrt{n}$-Discrimination of the Mean}

	We introduce the following notion of consistent $\sqrt{n}$-discrimination.

	\begin{definition}
		The mean of $X_{n,i}$, $n \ge 1$, is \bi{consistently $\sqrt{n}$-discriminated at level $\alpha \in (0,1)$ in model $\mathcal{P}_n$}, if there is a sequence of (potentially randomized) tests $\{\varphi_n\}_{n \ge 1}$ such that
		\begin{align*}
			\limsup_{n \rightarrow \infty} \mathbf{E}[\varphi_n(X_n)] \le \alpha,
		\end{align*}
		along any sequence $P_n \in \mathcal{P}_{n,0}$, and
		\begin{align*}
			\liminf_{n \rightarrow \infty}  \mathbf{E}[\varphi_n(X_n)] = 1,
		\end{align*}
		along any sequence $P_n \in \mathcal{P}_n$ such that $\lim_{n \rightarrow \infty} \sqrt{n} \mathbf{E}[X_{n,i}]/\sigma_{LR} = \infty$.
	\end{definition}\medskip

	We consider consistent discrimination against alternatives after normalizing by $\sigma_{LR}$ (which depends on $n$), so that when $\sigma_{LR}^2$ is larger, we focus on the alternative hypothesis that is farther away from the null hypothesis. Hence, if it is not possible to consistently $\sqrt{n}$-discriminate the mean, it is not necessarily due to the long run variance increasing to infinity fast.

     	The consistent $\sqrt{n}$-discrimination is often obtained when the parameter is in a finite dimensional space, and one knows the local dependence structure of the observations. To illustrate this point, suppose that $X_{n,1},...,X_{n,n}$'s are i.i.d. Then, often we have
     \begin{align*}
     	\frac{\sqrt{n}(\overline X_n - \mathbf{E}[X_{n,i}])}{\hat \sigma_n} \rightarrow_d \mathcal{N}(0,1),
     \end{align*}
     where $\hat \sigma_n^2 \rightarrow_p \sigma^2 = \Var(X_{n,1}) >0$, and $\overline X_n = \frac{1}{n}\sum_{i =1}^n X_{n,i}$. Let us consider the usual $t$-test as follows:
     \begin{align*}
     	\varphi_n(X_n) = 1\left\{ \frac{\sqrt{n}\overline X_n }{\hat \sigma_n}> z_{1-\alpha}\right\}.
     \end{align*}
     Under the Pitman local alternatives such that $\mathbf{E}[X_{n,i}] = \overline \mu/\sqrt{n}$, $\overline \mu>0$, we have
     \begin{align*}
     	\liminf_{n \rightarrow \infty}  \mathbf{E}[\varphi_n(X_n)] = 1 - \Phi\left( z_{1 - \alpha} - \overline \mu \right),
     \end{align*}
     where $\Phi$ denotes the CDF of $\mathcal{N}(0,1)$. The last term converges to 1 as $\overline \mu \rightarrow \infty$. Hence, the mean is consistently $\sqrt{n}$-discriminated. The discrimination results extend to the case with locally dependent observations where we know the local dependence structure and the long run variance is consistently estimable.

	 However, when we do not know the dependence structure, the consistent $\sqrt{n}$-discrimination of the mean is not guaranteed. We make this explicit in the following corollary which follows immediately from Theorem \ref{thm: consistent discrimination}.

     \begin{corollary}
		\label{cor: root n discr}
        Suppose that $\mathcal{P}_{n,\mathcal{N}} \subset \mathcal{P}_n$. Suppose further that $\alpha \in (0,1/2)$, and $M_n = 1$. Then, the consistent $\sqrt{n}$-discrimination of the mean at level $\alpha$ is not possible.
     \end{corollary}

	 On the other hand, if we have at least two large clusters and do not know the dependence structure within each cluster, we can consistently $\sqrt{n}$-discriminate the mean as long as the within-cluster sample means are asymptotically normal, as shown in the following theorem.

	\begin{theorem}
		\label{thm: root n discr}
		Suppose that there exists a sub-partition $\mathcal{M}_n' \subset \mathcal{M}_n$ such that $|\mathcal{M}_n'| \ge 2$ for each $n \ge 1$, and
		\begin{align}
			\liminf_{n \rightarrow \infty} \min_{ N_{n,m} \in \mathcal{M}_n'} \frac{|N_{n,m}|}{n} >0.
		\end{align}
        Suppose further that the set $\mathcal{M}_n'$ satisfies that for each $P_n \in \mathcal{P}_n$ and for each $t \in \mathbf{R}$,
	    \begin{align}
	    	\label{conv2}
	    	\max_{1\le m \le M_n : N_{n,m} \in \mathcal{M}_n'}\left| P_n\left\{ \frac{\sqrt{n_m} (\overline X_{m,n} - \mathbf{E}[X_{n,i}])}{\sigma_{n,m}} \le t \right\} - \Phi(t) \right| \rightarrow 0,
	    \end{align}
        as $n \rightarrow \infty$.

		Then, the mean is consistently $\sqrt{n}$-discriminated at level $\alpha \in (0,1)$.
	\end{theorem}

	For the theorem, we construct a $t$-test statistic as in \cite{Ibragimov/Muller:10:JBES} and show that  using the test, we can consistently $\sqrt{n}$-discriminate the mean, without knowing the dependence structure within the clusters.
	
	The asymptotic normality condition (\ref{conv2}) is often satisfied if the within-cluster dependence is weak. As we show later, this does \textit{not} mean that we can consistently estimate $\sigma_{n,m}$ for each cluster. (We will study this problem in the next section in detail.) Also, it is important to note that the within-cluster asymptotic normality is not enough to secure the consistent $\sqrt{n}$-discrimination of the mean, if there is only one cluster. In fact, the asymptotic normality condition alone does not exclude the possibility of $\mathcal{P}_{n,\mathcal{N}} \subset \mathcal{P}_n$, and in this case, Corollary \ref{cor: root n discr} shows that the mean is not consistently $\sqrt{n}$-discriminated. 

	As mentioned in the introduction, \cite{Song:16:arXiv} and \cite{Leung:21:JAE} considered the approach of randomized subsampling inference when one does not know the dependence structure at all. Hence, their situation corresponds to the setting with $M_n = 1$. Their procedure requires the following assumption: 
	\begin{align}
		\label{eq}
		\sqrt{n} \left( \overline X_{n} - \mathbf{E}[X_{n,i}] \right) = O_P(1),
	\end{align}
	as $n \rightarrow \infty$. If we know the upper bound of the long-run variance such that the upper bound does not change with $n$, it is not hard to see that we can consistently $\sqrt{n}$-discriminate the mean as long as the condition (\ref{conv2}) holds. Indeed, we can consider the test where we reject the null hypothesis of $\mathbf{E}[X_{n,i}] = 0$ against $\mathbf{E}[X_{n,i}] > 0$ if and only if 
	\begin{align*}
		\frac{\sqrt{n} \overline X_{n}}{\sqrt{c}} > z_{1 - \alpha},
	\end{align*}
	where $c$ is the known upper bound for the long-run variance. In our setting of hypothesis testing, however, the set of probabilities $\mathcal{P}_n$ does not have a finite upper bound for the long-run variance of the sample mean, reflecting the fact that the long-run variance is not known. Thus, the assumption (\ref{eq}) does not hold uniformly over $P \in \mathcal{P}_n$ in our setting, and the results of \cite{Song:16:arXiv} and \cite{Leung:21:JAE} do not contradict the impossibility result of Theorem \ref{thm: consistent discrimination}.\footnote{This setting is analogous to that in the standard hypothesis testing with i.i.d.\ normal random variables with the unknown variance. In this standard setting, even with the unknown variance, the mean is typically consistently $\sqrt{n}$-discriminated, because the variance can be consistently estimated. However, as we will see later, in a setting with large clusters, the long-run variance is not consistently estimable.}

	\section{Consistent Estimation of Variance}

	Recently, \cite{Hansen/Lee:19:JOE} showed that it is necessary and sufficient for the weak law of large numbers to hold for the sample average of the clustered observations with
	\begin{align*}
		\lim_{n \rightarrow \infty} \sum_{m = 1: n_m \ge 2}^{M_n} \left(\frac{n_m}{n}\right)^2 = 0.
	\end{align*}
    In this section, we show that this condition is necessary and sufficient for consistent estimability of the long run variance. This implies that when there is a large cluster (i.e., which takes up an asymptotically nonnegligible fraction of observations), the long run variance is not consistently estimable. This is a consequence of lack of knowledge of the dependence structure within the large cluster. It means that the usual asymptotic inference based on the asymptotic normal approximation of statistics is generally not applicable in this situation.

	\subsection{Consistent Estimability}

	Let us introduce the notion of consistent estimability of a parameter. Let $\mathcal{P}_n$ be the set of joint distributions of observed random variables, say, $\{X_1,...,X_n\}$. Given a parameter space $\Theta \subset \mathbf{R}^d$, we define our object of interest to be a map $\theta_n: \mathcal{P}_n \rightarrow \Theta$.

	\begin{definition}
		\label{def: consist estimability}
		For any sequence of subsets $\mathcal{P}_n' \subset \mathcal{P}_n$, we say that $\theta_n$ is \bi{consistently estimable in} $\mathcal{P}_n'$, if there exists an estimator $\hat \theta$ such that along any sequence $P_n \in \mathcal{P}_n'$,
		\begin{align*}
			P_n\left\{\| \hat \theta - \theta_n(P_n) \| > \epsilon \right\} \rightarrow 0,
		\end{align*}
	   as $n \rightarrow \infty$, for each $\epsilon>0$.
	\end{definition}

	One can find a similar definition of consistent estimability in \cite{LeCam/Schwartz:60:AMS}. They provide necessary and sufficient conditions for a parameter to be consistently estimable when the data are i.i.d. See also Section 1.4 of \cite{Ibragimov/Hasminskii:1981:StatEst} and Section 6.2 of \cite{Pfanzagl:94:ParamStatTheory}.

	Our setting is somewhat nonstandard, requiring a different technique to prove impossibility of consistent estimation. It is usually assumed that the probability model is indexed by a certain set, i.e., $\mathcal{P}_n = \{P_{n,h}: h \in \mathcal{H}\}$, where each $P_{n,h}$ is a probability measure indexed by $h$ in some topological space $\mathcal{H}$ that is independent of the sample size $n$. One can then redefine the parameter $\psi_n(h) = \theta_n(P_{n,h})$, $h \in \mathcal{H}$, i.e., as a map on $\mathcal{H}$. As long as $\psi_n$ behaves ``continuously'' on $\mathcal{H}$, the parameter $\psi$ can be shown to be consistently estimable. (See, e.g., Theorem 4.1 of \cite{Ibragimov/Hasminskii:1981:StatEst} and Theorem 6.2.11 of \cite{Pfanzagl:94:ParamStatTheory}.) Then the impossibility of consistent estimation stems from the discontinuity of $\psi$ as a map on $\mathcal{H}$, which yields ``non-identifiability'' of the parameter \citep{Potscher:02:Eca}.

    However, we cannot apply this standard approach in our setting, because there is no natural space $\mathcal{H}$ that indexes $\mathcal{P}_n$ independently of $n$. The main reason is that we need to deal with a situation potentially with a large cluster with an unknown within-cluster dependence structure. This means that we need to require our probability model to accommodate a wide range of dependence structures for the entire sample. For example, suppose that there is only one large cluster, so that one does not know the dependence structure at all. This means, among other things, that our model needs to include various network dependence structures \citep[such as those studied in][]{Kojevnikov/Marmer/Song:21:JOE} for the joint distribution of the entire random vector $[X_1,...,X_n]$ whose dimension grows with the sample size $n$. One might consider parametrizing the probabilities in terms of the networks governing the dependence structure, but each network depends on the sample size $n$. To the best of our knowledge, there is no obvious way to topologize such a probability model and to define the continuity of the parameter $\theta_n$ on the probabilities, independently of sample size $n$.

    Our approach relies on the following simple lemma that uses contiguity of probabilities at a primitive level. For any two sequences of probabilities $P_n$ and $P_n'$, we say that $P_n$ is \bi{contiguous with respect to} $P_n'$ if $P_n' (A_n) \rightarrow 0$ implies $P_n (A_n) \rightarrow 0$ for any sequence of Borel sets $A_n$, and write $P_n \triangleleft P_n'$. When $P_n \triangleleft P_n'$ and $P_n' \triangleleft P_n$, we say that $P_n$ and $P_n'$ are \bi{mutually contiguous}, and write $P_n \triangleleft \triangleright P_n'$. Contiguity between probabilities was introduced by \cite{LeCam:60:UCPS} and is widely used, especially for deriving the limiting distribution of a test statistic under local alternatives. By tracing out the limiting distribution along a range of local alternatives, one obtains a limiting experiment which one can use to compute the asymptotic risk lower bound in statistical decision theory. (See, e.g., Chapter 6 of \cite{vanderVaart:98:AsympStat}.)

	The following lemma summarizes our scheme of proving the impossibility of consistent estimability of $\sigma_{LR}^2$.
	\begin{lemma}
		\label{lemm: basic}
		Suppose that there exists a sequence $P_{n,0} \in \mathcal{P}_n$ such that $P_{n,1} \triangleleft P_{n,0}$ for every sequence $P_{n,1} \in \mathcal{P}_n$. Then, $\theta_n$ is consistently estimable in $\mathcal{P}_n$ if and only if $\lim_{n \rightarrow \infty}  \left\| \theta_n(P_{n,1}) - \theta_n(P_{n,0}) \right\| = 0$ for any sequence $P_{n,1} \in \mathcal{P}_n$.
	\end{lemma}
	Later we use Lemma \ref{lemm: basic} to prove the impossibility of consistent estimation of the long run variance, by selecting two Gaussian probabilities, $P_{n,0}$ and $P_{n,1}$, such that $P_{n,1} \triangleleft P_{n,0}$ and the values of the long run variance stay apart under $P_{n,0}$ and $P_{n,1}$ as $n \rightarrow \infty$. (See the discussion below Theorem \ref{thm: cluster dep}.)
	
	The notion of consistent estimability in Definition \ref{def: consist estimability} coincides with consistent estimability \textit{uniform in $P$}, i.e., the existence of an estimator $\hat \theta$ such that for each $\epsilon>0$, as $n \rightarrow \infty$,
	\begin{align*}
		\sup_{P_n \in \mathcal{P}_n}  P_{n}\left\{\| \hat \theta - \theta_n(P_{n}) \| > \epsilon \right\} \rightarrow 0.
	\end{align*}
	(See \cite{Ibragimov/Hasminskii:1981:StatEst}, p.31. See also \cite{Potscher:02:Eca} for discussion on asymptotics uniform in $P$.) When $\mathcal{P}_{n} = \{P_{n,h}: h \in \mathcal{H}\}$ for some index set $\mathcal{H}$ which does not depend on the sample size $n$, uniform consistent estimability is stronger than pointwise consistent estimability which assumes the existence of an estimator $\hat \theta$ such that for each $h \in \mathcal{H}$ and for each $\epsilon>0$,
		\begin{align*}
			P_{n,h}\left\{\| \hat \theta - \theta_n(P_{n,h}) \| > \epsilon \right\} \rightarrow 0,
		\end{align*}
	 as $n \rightarrow \infty$. However, as explained above, in our setting, there is no space $\mathcal{H}$ that indexes $\mathcal{P}_n$ and is independent of $n$. Hence, there is no natural notion of pointwise consistency in $P$ in our set-up.

\subsection{Consistent Estimability of Variance in Gaussian Experiments}
\subsubsection{A Necessary and Sufficient Condition for Consistent Estimability of Variance}
\label{subsec: Gaussian Experiments}

In our context, a major challenge is to show the contiguity condition (ii) of Lemma \ref{lemm: basic}. A standard argument proving contiguity utilizes the local asymptotic normality or local asymptotic mixed normality results for a log-likelihood process. However, these latter results often use an i.i.d.\ or time-series set-up where the researcher knows the dependence structure, and Hence, are not useful for our purpose here. For this reason, we focus on a Gaussian experiment, where we can explicitly compute the log-likelihood process in finite samples and investigate its asymptotic behavior as the dependence structure varies. In particular, we consider the following model for a fixed $\sigma^2>0$,
	\begin{eqnarray*}
		\mathcal{P}_{n,\mathcal{N}}(\sigma^2) = \left\{\Phi(\overline \mu \mathbf{1}_n,\Sigma_{n}(\sigma^2, \delta)): \delta \in (0,1] \text{ and }\overline \mu \in \mathbf{R} \right\},
	\end{eqnarray*}
where $\mathbf{1}_n$ denotes the $n$-dimensional vector of ones.  The set $\mathcal{P}_{n,\mathcal{N}}(\sigma^2)$ represents the set of LTIC Gaussian models, where each multivariate normal distributions with a common mean and the short run variance equal to $\sigma^2$.
    
For the impossibility result below, we require that the probability model does not exclude this Gaussian experiment.

\begin{theorem}
	\label{thm: cluster dep}
	Suppose that $\mathcal{P}_{n,\mathcal{N}}(\sigma^2) \subset \mathcal{P}_n$ for each $n \ge 1$, for some $\sigma^2>0$ that is independent of $n$. Then, the long run variance $\sigma_{LR}^2$ is consistently estimable in $\mathcal{P}_n$ if and only if
	\begin{align}
		\label{cond2}
        \lim_{n \rightarrow \infty} \sum_{m=1}^{M_n} \frac{n_m^2}{n^2} \rightarrow 0,
	\end{align}
    as $n \rightarrow \infty$.
\end{theorem}

The sufficiency part of the theorem is straightforward. To see this, suppose for simplicity that there is no singleton cluster in the data. If (\ref{cond2}) is satisfied, it means that the number of clusters $M_n$ grows to infinity as $n \rightarrow \infty$. Then, we consider the following estimator.
\begin{eqnarray*}
	\hat \sigma_{LR}^2 \equiv \frac{1}{n} \sum_{m=1}^{M_n} \sum_{i, j \in N_{n,m}} (X_{n,i} - \overline X_n)(X_{n,j} - \overline X_n).
\end{eqnarray*}
In fact, for the sufficiency part, we do not require that $\mathcal{P}_{n,\mathcal{N}}(\sigma^2) \subset \mathcal{P}_n$. 

The nontrivial part of the theorem is to show that the condition (\ref{cond2}) is \textit{necessary} for the consistent estimability of $\sigma_{LR}^2$ in $\mathcal{P}_{n,\mathcal{N}}$. Suppose that the condition of (\ref{cond2}) fails, which implies that one has at least one nonnegligible cluster. Then, we show that $\sigma_{LR}^2$ is not consistently estimable. For this, we employ Lemma \ref{lemm: basic} after computing the log-likelihood process under cluster dependence. More specifically, suppose first that the observations consist of only a single large cluster. Then, we note that the model $\mathcal{P}_n$, due to the lack of knowledge on the dependence structure, does not exclude the LTIC Gaussian experiment: $\Phi(0,\Sigma_n(\sigma^2,\delta))$. Then we show that
\begin{align*}
	\Phi(0,\Sigma_n(\sigma^2,0)) \triangleleft \Phi(0,\Sigma_n(\sigma^2,\delta)),
\end{align*}
whereas
\begin{align*}
	\sigma_{LR}^2\left(\Phi(0, \Sigma_n(\sigma^2,\delta))\right) - \sigma_{LR}^2\left(\Phi(0, \Sigma_n(\sigma^2,0))\right) \rightarrow c \ne 0,
\end{align*}
as $n \rightarrow \infty$, for some nonzero constant $c$. Hence, by Lemma \ref{lemm: basic}, $\sigma_{LR}^2$ cannot be consistently estimated in any probability model that does not exclude the LTIC Gaussian experiment. It is not hard to extend the same arguments to a setting where there are potentially multiple large clusters.

Theorem \ref{thm: cluster dep} then implies that if a nonnegligible portion of the sample belongs to non-singleton clusters, the long run variance $\sigma_{LR}^2$ is consistently estimable in $\mathcal{P}_{n,\mathcal{N}}$ if and only if the cluster structure consists of negligible clusters. We formalize this in the following corollary.

\begin{corollary}
	\label{cor: cluster dep2}
	Suppose that the conditions of Theorem \ref{thm: cluster dep} hold, and $\liminf_{n \rightarrow \infty} n^*/n > 0$, where $n^* = \sum_{m=1: n_m \ge 2}^{M_n} n_m$. Then, the long run variance $\sigma_{LR}^2$ is consistently estimable in $\mathcal{P}_n$ if and only if the cluster structure $\mathcal{M}_n$ consists of negligible clusters, i.e.,
	\begin{align}
		\label{cond32}
		\lim_{n \rightarrow \infty} \max_{1 \le m \le M_n} \frac{n_m}{n} = 0.
	\end{align}
\end{corollary}

The condition $\liminf_{n \rightarrow \infty} n^*/n > 0$ requires that the fraction of random variables $X_{n,i}$ that do not belong to a singleton cluster is asymptotically nonnegligible. In this case, if the probability model in practice includes the Gaussian model $\mathcal{P}_{n,\mathcal{N}}$ as a subclass and there is at least one nonnegligible cluster, it is not possible to consistently estimate the long run variance. Certainly, this impossibility result carries over to a model where the long run variance $\sigma_{LR}^2$ is allowed to increase with the sample size $n$.

Hence, by combining Theorem \ref{thm: root n discr} with Theorem \ref{thm: cluster dep}, we find that when we have several large clusters, the long run variance is not consistently estimable because the sample contains large clusters, but the mean can still be consistently $\sqrt{n}$-discriminated.

\subsubsection{Implications for Network Dependent Observations}

One might wonder whether the result extends to the case where the observations exhibit a dependence structure other than cluster dependence.  Below we give a partial answer for the case of a dependency graph. Dependency graphs were introduced by \cite{Stein:86}, and have been studied and used in statistics and econometrics. (See, e.g., \cite{Aronow&Samii:17}, \cite{Song:18:ReStat}, \cite{Leung:20:ReStat} and \cite{Canen/Schwartz/Song:20:QE} and references therein.)

A \bi{graph} (or network) is a pair $G_n = (N_n,E_n)$, where $N_n = \{1,...,n\}$ denotes the set of vertices and $E_n$ the set of edges, where we denote $N(i) = \{j: ij \in E_n\}$ to mean the neighborhood of vertex $i$. (Here, we consider only simple, undirected graphs, i.e., $ii \notin E_n$, for all $ i \in N_n$, and $ij \in E_n$ if and only if $ji \in E_n$.) We define
\begin{align*}
	d_{mx} = \max_{i \in N_n} |N(i)| \quad \text{ and } \quad d_{av} = \frac{1}{n}\sum_{i \in N_n} |N(i)|,
\end{align*}
where $d_{mx}$ is called the \bi{maximum degree}, and $d_{av}$ the \bi{average degree} of the graph $G_n$. The maximum and average degrees are often used to capture the denseness of the graph. A subset of vertices in graph $G_n$ is called a \bi{clique} if any two distinct vertices are adjacent in $G_n$, and the number of vertices in the clique is called the size of the clique. The maximum clique size refers to the size of the clique that is largest in the graph $G_n$.

 Recall that a graph $G_n = (N_n,E_n)$ on $N_n = \{1,...,n\}$ is called a \bi{dependency graph} for $X_n = (X_{n,i})_{i \in N_n}$, if for any subset $A \subset N_n$, $(X_{n,i})_{i \in A}$ and $(X_{n,i})_{i \in N_n \setminus \overline N_n(A)}$ is independent, where $\overline N_n(A) = \{j: ij \in E_n, \text{ for some } i \in A\} \cup \{i\}$. It is important to note that while the dependency graph imposes independence between $X_{n,i}$ and $X_{n,j}$ when they are not adjacent in the graph, it says nothing about dependence between them when they are adjacent. Thus, we allow in $\mathcal{P}_n$ any degree of dependence (including independence) between $X_{n,i}$ and $X_{n,j}$ whenever $i$ and $j$ are adjacent in $G_n$. As the dependency graph becomes denser, this reflects our limited knowledge on the dependence structure, similarly to large clusters in the cluster dependence case.

\begin{corollary}
	\label{cor: dep graph}
	Suppose that the conditions of Theorem \ref{thm: cluster dep} hold and that for each $n \ge 1$, there exists a graph $G_n = (N_n,E_n)$ which has maximum degree $d_{mx}$, average degree $d_{av}$, maximum clique size $n_C$, and each distribution $P_n \in \mathcal{P}_n$ of $X_n$ has $G_n$ as a dependency graph. Then, the following holds.\medskip

	(i) If $\limsup_{n \rightarrow \infty} n_C/ n > 0$, the long run variance $\sigma_{LR}^2$ is not consistently estimable.

	(ii) If $\lim_{n \rightarrow \infty} d_{mx}^2 d_{av} /n = 0$, the long run variance $\sigma_{LR}^2$ is consistently estimable.
\end{corollary}

The impossibility result in (i) has an important implication in many models with network dependent observations. As in the case of a dependency graph, many models of network dependence do not specify the strength of dependence between observations that are adjacent in the network \citep[e.g.,][]{Kojevnikov/Marmer/Song:21:JOE}. Weak dependence is usually imposed between observations that are far from each other in terms of the shortest path in the network. Hence, when there is a large clique in the network which constitutes a nonnegligible fraction of the entire sample in the limit as $n \rightarrow \infty$, Corollary \ref{cor: dep graph}(i) implies that the long run variance of the network dependent observations is not consistently estimable.

It is interesting to note that one cannot characterize a necessary and sufficient condition for the network solely in terms of its maximum degree. For example, if $X_n$ is a multivariate normal random vector such that each component has a bounded variance and has a star graph as a dependency graph, the long run variance is consistently estimable. To see this, let $X_n = [X_{n,1},...,X_{n,n}]^\top$ be a centered multivariate random vector which has a graph $G_n$ as a dependency graph. Let the graph $G_n$ be a star graph with the unit $1$ being its center.\footnote{The star graph as a dependency graph is different from an additive common shock model such as $X_{n,i} = C_n + \varepsilon_i$, where $C_n$ is a common shock, and $\varepsilon_i$'s are cross-sectionally independent idiosyncratic shock. In this case, the dependency graph is a \textit{complete graph}, because every pair of random variables is correlated through the common shock. Hence, the center in the star graph as a dependency graph cannot be a source like a common shock. It is more plausible to imagine the center to be an aggregated outcome of independent sources. In this case, by simply eliminating the star, one obtains independent random variables.} In the context of multivariate normality, we can write
\begin{eqnarray*}
	X_{n,1} = \sum_{i \in N_n \setminus \{1\}} \theta_i X_{n,i} + \varepsilon,
\end{eqnarray*}
where the leading sum is the best linear projection, so that $\varepsilon$ is independent of $X_{n,i}$'s, $i \in N_n\setminus\{1\}$, which are independent from each other (due to the dependency graph being a star graph). Since the variance of $X_{n,1}$ is bounded, we should have $\sum_{i \in N_n \setminus \{1\}} \theta_i^2 < C$, for all $n \ge 1$, for some $C>0$. Note that we can identify
\begin{eqnarray*}
	\theta_i = \Cov(X_{n,i}, X_{n,1})/\Var(X_{n,i}).
\end{eqnarray*}
Now, we can write
\begin{align*}
	\sigma_{LR}^2 = \frac{1}{n} \sum_{i=1}^n \mathbf{E}[X_{n,i}^2] + \frac{1}{n} \sum_{i=1}^n \sum_{j=1, i \ne j}^n \mathbf{E}[X_{n,i} X_{n,j}].
\end{align*}
The second term is written as
\begin{align*}
	 \frac{2}{n} \sum_{i=1, i \ne 1}^n \mathbf{E}[X_{n,i} X_{n,1}] = 2\mathbf{E}\left[\left( \frac{1}{n}\sum_{i=1,i\ne 1}^n X_{n,i} \right) X_{n,1}\right] = o(1),
\end{align*}
as $n \rightarrow \infty$, because the normalized sum in the parenthesis converges to zero in moments. Hence, we can simply take
\begin{eqnarray*}
	\hat \sigma_{LR}^2 = \frac{1}{n}\sum_{i =1}^n X_{n,i}^2
\end{eqnarray*}
to be an estimator of the long run variance. It is not hard to see that $\hat \sigma_{LR}^2$ is consistent for $\sigma_{LR}^2$. This example shows that one cannot express the condition for the consistent estimability solely in terms of the maximum degree of the dependency graph.

\section{Implications}
\subsection{A Linear Regression Model with Cluster-Dependent Errors}
Let us consider the following regression model with cluster-dependent errors (see, e.g., \cite{Cameron/Gelbach/Miller:08:ReStat}, \cite{Djogbenou/MacKinnon/Nielsen:19:JOE} and \cite{Hansen/Lee:19:JOE} and references therein):
\begin{align*}
	y = X \beta + u,
\end{align*}
where $y = [y_1',...,y_{M_n}']'$, $X = [X_1',...,X_{M_n}']'$ and $u = [u_1',...,u_{M_n}]'$, with $\mathbf{E}[u_m \mid X] = 0$ for each $m=1,...,M_n$, and each cluster $m$ has $n_m$ observations (so that $y_m$ and $u_m$ are $n_m$ dimensional column vectors, and $X_m$ is an $n_m \times k$ matrix.) We assume that $u_1,...,u_{M_n}$ are independent, but for each $m$, the dependence structure of $u_m$ is not known. We do not exclude the possibility that the error term follows a normal distribution. 

Then, the OLS estimator of $\beta$ is given by
\begin{align*}
	\hat \beta = (X'X)^{-1}X' y.
\end{align*}
The sandwich form of the variance matrix of $\hat \beta$ is given by
\begin{align*}
	V = (X'X)^{-1} \left(\sum_{m=1}^{M_n} X_m' \mathbf{E}\left[u_m u_m' \mid X \right] X_m \right) (X'X)^{-1}.
\end{align*}
Once we obtain a consistent estimator $\hat V$ of $V$, we can construct a standard error of the $j$-th entry of $\beta$, i.e., $\beta_j$, as $\hat \sigma_j^2 = [\hat V]_{jj}$, the $j$-th diagonal of $\hat V$. From the asymptotic normal inference applied to a $t$-statistic for $\beta_j$, we obtain the following confidence interval for $\beta_j$: 
\begin{align}
	\label{CI}
	\left[\hat \beta_j - \frac{z_{\alpha/2} \hat \sigma_j}{\sqrt{n}}, \hat \beta_j + \frac{z_{\alpha/2} \hat \sigma_j}{\sqrt{n}}\right].
\end{align}
As for the consistent estimator $\hat V$, \cite{Djogbenou/MacKinnon/Nielsen:19:JOE} considered the following estimator:
\begin{align*}
	\hat V = d (X'X)^{-1} \left(\sum_{m=1}^{M_n} X_m' \hat u_m \hat u_m' X_m\right) (X'X)^{-1},
\end{align*}
where $\hat u_m = y_m - X_m' \hat \beta$, and $d$ is a sequence such that $d \rightarrow 1$. They established the consistency of this estimator under a set of conditions, and showed that their conditions are not compatible with a setting in which one of the clusters is large, i.e., its size is proportional to the entire sample.

Our result implies that such an estimator $\hat V$ is not uniformly consistent when there is at least one large cluster. In fact, our result is much stronger than this. It shows that it is not possible to construct a uniformly consistent estimator of $V$ in such a case. Hence, in this case, we cannot construct a confidence interval of the form (\ref{CI}) that is uniformly asymptotically valid. When a nonnegligible fraction of observations belong to a (non-singleton) cluster - which is the case with most cluster-dependence settings, the necessary and sufficient condition for the uniformly consistent estimability of $V$ is that each cluster is asymptotically negligible in the sense of (\ref{cond32}).

\subsection{Difference-in-Differences with Spillovers}
Let us explore the implications of the impossibility results in the context of a difference-in-differences approach to causal inference. (See Section 6.5 of \cite{Imbens/Wooldridge:09:JEL} for an overview of this method. See also \cite{Roth/SantAnna/Bilinski/Poe:22:arXiv} for an overview including recent advances in the literature.) Suppose that there are $n$ individuals who are subject to a treatment and the researcher observes their outcomes before and after the treatment. We let $Y_{i,t}(1)$ and $Y_{i,t}(0)$
denote the potential outcomes at time $t = 0,1$ for the treated state and the control state, respectively. As standard in the literature, we assume that in time 0, no individual is treated, and $Y_{i,0} = Y_{i,0}(0)$, which is observed. The observed outcome $Y_{i,1}$ at time $1$ is defined by
\begin{eqnarray*}
	Y_{i,1} = D_i Y_{i,1}(1) + (1 - D_i) Y_{i,1}(0),
\end{eqnarray*}
where $D_i$ is the indicator of treatment for $i$ that happens between times 0 and 1. Our parameter of interest is the average treatment effect on the treated:
\begin{eqnarray*}
	\text{ATT} = \mathbf{E}\left[ Y_{i,1}(1)-Y_{i,1}(0)\mid D_i = 1\right].
\end{eqnarray*}
Suppose that we have observations $\{(Y_{i,1},Y_{i,0},D_i)\}_{i=1}^n$, where $Y_{i,0}$ is the outcome for person $i$ at time 0. Furthermore, we assume that the researcher knows the probability $p = P\{D_i =1\}$. (The impossibility result we mention below carries over to the case where $p$ is not known.)

Let us introduce the standard parallel trend assumption used in the literature:
\begin{align*}
	\mathbf{E}\left[ Y_{i,1}(1) - Y_{i,1}(0) \mid D_{i}=1\right] =\mathbf{E}\left[ Y_{i,1}(1) - Y_{i,1}(0) \mid D_{i}=0\right].
\end{align*}
Under this assumption, we can identify
\begin{align*}
	\text{ATT} = \mathbf{E}\left[ \Delta Y_{i}\mid D_{i}=1\right] -\mathbf{E}\left[\Delta Y_{i}\mid D_{i}=0\right],
\end{align*}
where $\Delta Y_{i} = Y_{i,1} - Y_{i,0}$. We can obtain a sample analogue estimator by 
\begin{align*}
	\widehat{\text{ATT}} = \frac{1}{n}\sum_{i=1}^n \frac{D_i \Delta Y_{i}}{p} - \frac{1}{n}\sum_{i=1}^n \frac{(1 - D_i) \Delta Y_{i}}{1-p}.
\end{align*}
We consider settings where the observed outcomes are cross-sectionally dependent. Our interest is in constructing a confidence interval for $\text{ATT}$ that is uniformly asymptotically valid. Below we consider two situations, one with treatment spillover and the other with spillover of treatment effects. We explore implications of our impossibility results in these situations.

\subsubsection{Treatment Spillover}

Suppose that there is a spillover of the treatments so that $D_i$'s are correlated across $i$, along some network among people. For example, one can think of a situation in a social program where two people $i$ and $j$ are neighbors and participating in the program by $i$ can induce the participation by $j$. Suppose that the researcher does not have information on the neighborhoods among the subjects. Then, this creates dependence among $Y_{i,1}$'s along a dependence structure that is unknown to the researcher. Then, our impossibility result shows that $\text{ATT}$ cannot be consistently discriminated.

In practice, the treatment assignment is often done at the cluster level, where the potential outcomes $Y_{i,1}(1)$ and $Y_{i,1}(0)$ may exhibit  arbitrary dependence within each cluster. (See Section 5 of \cite{Roth/SantAnna/Bilinski/Poe:22:arXiv} for examples and references studying such a setting.) 

Suppose that we have at least two large clusters such that $(Y_{i,1}(1), Y_{i,0}(0), D_i)$ are independent across the clusters but arbitrarily correlated within each cluster. The researcher might attempt to test the null hypothesis of $\text{ATT} = 0$ by considering the usual $t$ statistic for testing the null hypothesis of $\text{ATT} = 0$ such that
\begin{align*}
	t = \frac{\sqrt{n}(\widehat{\text{ATT}} - \text{ATT})}{\widehat{\sigma}},
\end{align*}
where $\widehat{\sigma}^2$ is a consistent estimator of the variance of $\sqrt{n}(\widehat{\text{ATT}} - \text{ATT})$, and the critical values taken from the standard normal stable. Our impossibility result implies that it is not possible to consistently estimate the variance of $\sqrt{n}(\widehat{\text{ATT}} - \text{ATT})$, when there is at least one large cluster, and hence, such a $t$-test is not uniformly asymptotically valid. For the same reason, we cannot construct a confidence interval of the following familiar form: 
\begin{align}
	\label{CI2}
	\left[\widehat{\text{ATT}} - \frac{z_{\alpha/2} \widehat{\sigma}}{\sqrt{n}}, \widehat{\text{ATT}} + \frac{z_{\alpha/2} \widehat{\sigma}}{\sqrt{n}}\right]
\end{align}
such that the confidence interval is uniformly asymptotically valid. (See Section 5 of \cite{Roth/SantAnna/Bilinski/Poe:22:arXiv} for various approaches.\footnote{To the best of our knowledge, there is no formal result that proposes a uniformly asymptotically valid confidence interval for $\text{ATT}$ in this setting. However, we expect that the bootstrap approach of \cite{Canay/Santos/Shaikh:21:ReStat} can be used to construct a uniformly valid confidence interval under mild additional conditions.})

\subsubsection{Spillover of Treatment Effects}

Suppose that the treatments $D_i$ themselves do not exhibit any spillover, but the cross-sectional dependence of $(Y_{i,1}(1), Y_{i,0}(0))$ arises due to the spillover of the treatment effects, for example, the treatment of a person $i$ influences the outcome of the person $j$ in the next period. Such a setting has been studied in the recent literature (see \cite{Aronow&Samii:17}, \cite{Leung:20:ReStat}, \cite{He/Song:22:WP} and references therein.)

Suppose that the spillover of the treatment effects arises along some network among people, and yet the researcher does not have any information on the network. Then, our impossibility result implies that we cannot consistently discriminate ATT in such a situation. However, the researcher may observe a group structure where the spillover does not arise between groups, so that $(Y_{i,1}(1), Y_{i,0}(0), D_i)$ are independent across groups.

If each within-group sum of $(Y_{i,1}(1), Y_{i,0}(0), D_i)$ satisfies the central limit theorem, our result shows that the ATT can be consistently $\sqrt{n}$-discriminated. However, when there is at least one large group, there does not exist a consistent estimator of the variance of $\sqrt{n}(\widehat{\text{ATT}} - \text{ATT})$. Hence, similarly as before, we cannot construct a uniformly asymptotically valid $t$-test for the null hypothesis of $\text{ATT} = 0$ using the usual $t$ statistic and standard normal critical values, and cannot construct a uniformly asymptotically valid confidence interval of the form (\ref{CI2}) based on a normal approximation.

\section{Conclusion}

In this paper, we show two impossibility results on the inference on the mean when the dependence structure is not known. The first result is the impossibility of consistent estimation of the long run variance. The second result is the impossibility of the consistent $\sqrt{n}$-discrimination of the mean. We made an attempt to accommodate partial knowledge of the dependence structure through cluster dependence, and has obtained some necessary and sufficient conditions for the cluster structure for the impossibility results.

While cluster dependence is a popularly used specification of the cross-sectional dependence structure, it is not general enough to accommodate other forms of a dependence structure such as dependency graphs, Markov graphs, and network dependence. It would be interesting to investigate the implications of partial knowledge of a dependence structure for a more general setting. We leave this for future research.

\section{Appendix: Mathematical Proofs}

\subsection{Preliminary Results}

For the proof of the main results, we first prove auxiliary lemmas. As a first step, we provide an explicit form of a log-likelihood process in Gaussian experiments in Lemma \ref{lemma: log LR}. For this, we use the following auxiliary lemma.

\begin{lemma}
	\label{lemm: aux 1}
	Let $\Sigma_0 = U S U^\top$ be the spectral decomposition of an $n \times n$, symmetric positive definite matrix $\Sigma_0$ and let $\Sigma_1$ be an $n \times n$ matrix defined as
	\begin{eqnarray*}
		\Sigma_1 = \Sigma_0 + U A U^\top
	\end{eqnarray*}
	for some symmetric positive semidefinite matrix $A$. Let $B \Lambda B^\top$ be the spectral decomposition of $S^{-1/2} A S^{-1/2}$. Suppose that $|\lambda_i| < 1$ for all $i=1,...,n$, where $\lambda_i$ denote the $i$-th diagonal entry of $\Lambda$.

	Then the following results hold.

	(i)
	\begin{eqnarray}
		\log\left(|\Sigma_1|^{-1/2}\right) - \log\left(|\Sigma_0|^{-1/2}\right) = - \frac{1}{2}\sum_{i=1}^n \log\left( 1 + \lambda_i \right).
	\end{eqnarray}

	(ii) For any vectors $a,b \in \mathbf{R}^n$,
	\begin{eqnarray}
		a^\top(\Sigma_1^{-1} - \Sigma_0^{-1})b = -\sum_{i=1}^n \frac{\lambda_i}{1 + \lambda_i} \tilde a_i \tilde b_i,
	\end{eqnarray}
	where $\tilde a_i$ and $\tilde b_i$ are the $i$-th entries of $\tilde a$ and $\tilde b$, with
	\begin{eqnarray}
		\tilde a = B^\top S^{-1/2} U^\top a \quad\text{and}\quad \tilde b = B^\top S^{-1/2} U^\top b.
	\end{eqnarray}
\end{lemma}

\noindent \textbf{Proof: } Let $Q=U S^{1/2}$ such that $\Sigma_0 =Q B B^{\top}Q^{\top}$ and $\Sigma_{n,1}=Q B(I +\Lambda) B^{\top} Q^{\top}$. Thus,
\begin{align*}
	|\Sigma_1|=|\Sigma_0|\cdot| I +\Lambda|=|\Sigma_0|\cdot\prod_{i=1}^n(1+\lambda_i),
\end{align*}
and
\begin{align*}
	\Sigma_0^{-1}-\Sigma_1^{-1} =(B^{\top}Q^{\top})^{-1}\left(I-(I+\Lambda)^{-1}\right)(Q B)^{-1}
	=U S^{-1/2}B \tilde{\Lambda} B^{\top} S^{-1/2} U^{\top},
\end{align*}
where
\[
\tilde{\Lambda} = \diag\left(\frac{\lambda_1}{1+\lambda_1},\ldots,\frac{\lambda_n}{1+\lambda_n}\right).
\]
$\blacksquare$\medskip

The following lemma provides an explicit form of a general log-likelihood process for Gaussian measures. Recall that $\Phi(\mu,\Sigma)$ denotes the multivariate normal distribution with mean vector $\mu$ and covariance matrix $\Sigma$.

\begin{lemma}
	\label{lemma: log LR}
	Let $\Sigma_0$, $\Sigma_1, \Lambda, B, S$ and $U$ be the matrices in Lemma \ref{lemm: aux 1}. Then, for all $x, \mu_1, \mu_0 \in \mathbf{R}^n$,
	\begin{align*}
		\quad \quad \log \frac{d \Phi(\mu_1,\Sigma_1)}{d \Phi(\mu_0,\Sigma_0)}(x) = -  \sum_{i=1}^n \log q_{i} + \frac{1}{2}\sum_{i=1}^n \frac{1}{q_i^2}\left(Z_i(x) (q_i + 1) - \tilde \mu_i \right)\left(Z_i(x) (q_i -1) + \tilde \mu_i \right),
	\end{align*}
	where $q_i = \sqrt{1 +  \lambda_i}$, $\lambda_i$ is the $i$-th diagonal entry of $\Lambda$, $Z_i(x)$ is the $i$-th entry of $Z(x)$ and $\tilde \mu_i$ is the $i$-th entry of $\tilde \mu$ with
	\begin{eqnarray}
		Z(x) = B^\top S^{-1/2} U^\top(x - \mu_0) \quad\text{and}\quad
		\tilde \mu = B^\top S^{-1/2} U^\top (\mu_1 - \mu_0).
	\end{eqnarray}
\end{lemma}

\noindent \textbf{Proof:} We write
\begin{align}
	\label{Log Lik}
	\begin{aligned}
		& \log \frac{d \Phi(\mu_1,\Sigma_1)}{d \Phi(\mu_0,\Sigma_0)}(x)
		= \log\left(|\Sigma_1|^{-1/2}\right) - \log\left(|\Sigma_0|^{-1/2}\right) \\
		& \qquad  - \frac{1}{2}\left( \left(x - \mu_1\right)^\top \Sigma_1^{-1} \left(x - \mu_1\right) - \left(x - \mu_0\right)^\top\Sigma_0^{-1} \left(x - \mu_0\right) \right).
	\end{aligned}
\end{align}
We apply Lemma \ref{lemm: aux 1}(i) to the first term on the right hand side. As for the last term we let $\mu_{\Delta} = \mu_1 - \mu_0$, and $x_* = x - \mu_0$. Note that
\begin{align*}
	\mu_{\Delta}^\top \Sigma_0^{-1} \mu_{\Delta} &= \mu_{\Delta}^\top U S^{-1} U \mu_{\Delta}
	=\mu_{\Delta}^\top U S^{-1/2} S^{-1/2} U^\top \mu_{\Delta}\\
	&=\mu_{\Delta}^\top U S^{-1/2} B B^{\top} S^{-1/2} U^\top \mu_{\Delta} = \tilde \mu^\top \tilde \mu.
\end{align*}
Similarly, $\mu_{\Delta}^\top \Sigma_0^{-1} x_* =  \tilde \mu^\top Z(x).$ We rewrite the last term in (\ref{Log Lik}) as
\begin{align*}
	& - \frac{1}{2}x_*^\top (\Sigma_1^{-1} - \Sigma_0^{-1}) x_*
	- \frac{1}{2} \left(\mu_{\Delta}^\top (\Sigma_1^{-1} - \Sigma_0^{-1}) \mu_{\Delta} - 2\mu_{\Delta}^\top (\Sigma_1^{-1} - \Sigma_0^{-1}) x_* \right) \\
	&\qquad \quad  \quad - \frac{1}{2} \left(\mu_{\Delta}^\top\Sigma_{0}^{-1}\mu_{\Delta} - 2 \mu_{\Delta}^\top\Sigma_{0}^{-1} x_* \right)\\
	&\qquad= \frac{1}{2} \sum_{i=1}^n \frac{ \lambda_i}{1 +  \lambda_i} Z_i^2(x)
	+\frac{1}{2} \sum_{i=1}^n \frac{ \lambda_i}{1 +  \lambda_i}\left(\tilde \mu_i^2 - 2 \tilde \mu_i Z_i(x) \right)- \frac{1}{2} \sum_{i=1}^n \left(\tilde \mu_i^2 - 2 \tilde \mu_i Z_i(x)\right)\\
	&\qquad= \frac{1}{2} \sum_{i=1}^n \left(\left(\frac{ \lambda_i}{1 +  \lambda_i} - 1\right) \left(Z_i(x) - \tilde \mu_i \right)^2 + Z_i^2(x) \right),
\end{align*}
(by applying Lemma \ref{lemm: aux 1}(ii)). By rearranging terms, we rewrite the last sum as
\begin{eqnarray}
	\frac{1}{2} \sum_{i=1}^n \frac{1}{q_i^2}\left(Z_i(x) (q_i+1) - \tilde \mu_i \right)\left(Z_i(x) (q_i-1) + \tilde \mu_i \right).
\end{eqnarray}
Combining this with an earlier result, we obtain the desired result. $\blacksquare$

\begin{lemma}
	\label{lemma: LR for the mean}
	Let $\Sigma_n = \sigma^2\left( (1 - \delta) I_n + \delta \mathbf{1}_n \mathbf{1}_n^\top \right)$, where $\delta$ is such that $n\delta \in (-1,1)$.

	Then, for any $\overline \mu \in \mathbf{R}$ and $x \in \mathbf{R}^n$, we have
	\begin{align*}
		\log \frac{d \Phi(\overline \mu \mathbf{1}_n,\Sigma_n)}{d \Phi(0,I_n)}(x) &= - \log \sqrt{\sigma^2(1 + (n-1) \delta)} - (n-1) \log \sqrt{\sigma^2(1 - \delta)}\\
		& \quad + \frac{\sigma^2 (1 + (n-1) \delta) - 1}{2\sigma^2 (1 + (n-1) \delta)} Z_1^2(x) + \frac{\sigma^2 (1- \delta) - 1}{2 \sigma^2 (1 - \delta)} \sum_{k=2}^n Z_k^2(x)\\
		& \quad + \frac{\sqrt{n}\overline \mu}{\sigma^2(1 + (n-1)\delta)} Z_1(x) - \frac{n \overline \mu^2}{2\sigma^2 (1 + (n-1)\delta)},
	\end{align*}
	where $Z_k(x)$ is the $k$-th entry of $Z(x) \equiv B^\top x$, and $B = [b_1,...,b_n]$ is an $n \times n$ orthogonal matrix such that $b_1 = n^{-1/2} \mathbf{1}$, $b_k^\top \mathbf{1} = 0$ for all $k=2,...,n$, $b_k^\top b_\ell = 0$ for all $k \ne \ell = 2,...,n$.
\end{lemma}

\noindent \textbf{Proof: } We apply Lemma \ref{lemma: log LR} with $S = U = I_n$, 
\begin{align*}
    A = \left( \sigma^2(1 - \delta) - 1 \right) I_n + \sigma^2 \delta \mathbf{1}_n \mathbf{1}_n^\top.
\end{align*}
Note that the spectral decomposition of $A$ is given by $B \Lambda B^\top$, where $\Lambda$ is the diagonal matrix with the diagonal elements $\lambda_1,...,\lambda_n$ given as $\lambda_1 = (\sigma^2 - 1) + \sigma^2(n-1)\delta$, $\lambda_2 = ... = \lambda_n = - \sigma^2 \delta$, and the orthogonal matrix $B$ as given the lemma. The desired result follows from Lemma \ref{lemma: log LR}. $\blacksquare$\medskip

Lemma \ref{lemma: LR for the mean} yields the following result for the case with cluster dependence. From here on, we make the dimension of the matrices and vectors explicit. Let $I_{n_m}$ be the $n_m$-dimensional identity matrix and $\mathbf{1}_{n_m}$ denote the $n_m$-dimensional column vector of ones.

\begin{corollary}
	\label{cor: log LR cluster}
	Let $\Sigma_n$ be the block diagonal matrix whose $m$-th block, $m = 1,...,M_n$, is given by
	\begin{eqnarray}
		\label{Sigma_nm}
		\Sigma_{n,m} = \sigma^2 \left((1 - \delta_{n,m}) I_{n_m} +  \delta_{n,m} \mathbf{1}_{n_m} \mathbf{1}_{n_m}^\top \right),
	\end{eqnarray}
	for some $\delta_{n,m} \in \mathbf{R}$ such that $n_m \delta_{n,m} \in (-1,1)$.

	Then, for any $\overline \mu_n \in \mathbf{R}$, and $x = [x_1,...,x_n]' \in \mathbf{R}^n$, 
	\begin{align*}
		\log \frac{d \Phi(\overline \mu_n \mathbf{1}_n,\Sigma_n)}{d \Phi(0,I_n)}(x) &= - \sum_{m=1}^{M_n} \log \sqrt{\sigma^2(1 + (n_m-1) \delta_{n,m})} - \sum_{m=1}^{M_n} (n_m-1) \log \sqrt{\sigma^2(1 - \delta_{n,m})}\\
		& \quad + \sum_{m=1}^{M_n} \frac{\sigma^2 (1 + (n_m-1) \delta_{n,m})-1}{2\sigma^2 (1 + (n_m-1) \delta_{n,m})}Z_{m,i_m}^2(x) \\
		& \quad + \sum_{m=1}^{M_n}  \frac{\sigma^2 (1-\delta_{n,m}) - 1}{2 \sigma^2(1 -  \delta_{n,m})} \sum_{i \in N_{n,m} \setminus \{i_m\}} Z_{m,i}^2(x)\\
		& \quad + \sum_{m=1}^{M_n} \frac{n\overline \mu_n/\sqrt{n_m}}{\sigma^2 (1 + (n_m-1)\delta_{n,m})} Z_{n,i_m}(x) - \frac{1}{2}\sum_{m=1}^{M_n} \frac{n^2\overline \mu_n^2/\sqrt{n_m}}{\sigma^2(1 + (n_m - 1)\delta_{n,m})},
	\end{align*}
	where $Z_{m,i}(x)$, $i \in N_{n,m}$, are the entries of $Z_{m}(x) = B_m^\top x_{m,n}$, $B_m$ is an $n_m \times n_m$ orthogonal matrix, $i_m$ denotes the first index in $N_{n,m}$, and $x_{m,n} = [x_i]_{i \in N_{n,m}}$.
\end{corollary}

\begin{lemma}
	\label{lemm: bound basic}
	Suppose that $f: \mathbf{R} \rightarrow \mathbf{R}^{+}$ is a continuously differentiable function such that for some $\delta >0$,
	\begin{eqnarray*}
		\sup_{x \in [-\delta, \delta]} \left| \frac{d \log f(x)}{dx}\right| |\overline x| < 1,
	\end{eqnarray*}
	where $\overline x \in \mathbf{R}$ is such that
	\begin{eqnarray*}
		\sup_{x \in [-\delta,\delta]} f(x) = f(\overline x).
	\end{eqnarray*}

	Then,
	\begin{eqnarray*}
		f(\overline x) \le \left( 1 - \sup_{x \in [-\delta, \delta]}\left| \frac{d \log f(x)}{dx}\right| | \overline x | \right)^{-1} f(0).
	\end{eqnarray*}
\end{lemma}

\noindent \textbf{Proof: } Using the Mean Value Theorem,
\begin{align*}
	\exp \log f(x) &= \exp \log f(0) + \frac{d \log f(x^*(x))}{dx} \left(\exp \log f(x^*(x)) \right) x\\
	&\le \exp \log f(0) + \sup_{\tilde x \in [-\delta, \delta]} \left| \frac{d \log f(\tilde x)}{dx} \right| \left(\exp \log f(\overline x)\right) |x|,
\end{align*}
where $x^*(x)$ is a point on the line segment between $0$ and $x$. Evaluating the inequality at $x=\overline x$ gives us the desired result. $\blacksquare$ \medskip

Recall the defintion of $n^*$ in Corollary \ref{cor: cluster dep2}:
\begin{eqnarray}
	\label{n star}
	n^* = \sum_{m=1: n_m \ge 2}^{M_n} n_m.
\end{eqnarray}
The number $n - n^*$ represents the number of random variables, $X_{n,i}$, that are known to be mutually independent. Each variable outside this set belongs to a non-singleton cluster.

\begin{lemma}
	\label{lemm: equiv}
	$\lim_{n \rightarrow \infty} \sum_{m = 1}^{M_n} \left(n_m / n\right)^2 = 0$ if and only if

	(a) $\lim_{n \rightarrow \infty} n^*/n = 0$, or

	(b) $\lim_{n \rightarrow \infty} \sum_{m = 1: n_m \ge 2}^{M_n} \left(n_m / n^*\right)^2 = 0$.
\end{lemma}

\noindent \textbf{Proof: } For each $n \ge 1$, we have either
\begin{align*}
	\sum_{m=1}^{M_n} \frac{n_m^2}{n^2} = \left\{\begin{array}{ll}
		                      \displaystyle \frac{1}{n}, & \text{ if } n^* =0,\\
		                      \displaystyle \left(\frac{n^*}{n}\right)^2 \sum_{m=1: n_m \ge 2}^{M_n} \left(\frac{n_m}{n^*}\right)^2 + \left( \frac{1}{n}\right)^2 (n - n^*), & \text{ if } n^* >0.
		                    \end{array}
	                    \right.
\end{align*}
Since $(1/n)^2(n- n^*) = o(1)$, $\lim_{n \rightarrow \infty} \sum_{m = 1}^{M_n} \left(n_m / n\right)^2 = 0$ if and only if (a) or (b) holds. $\blacksquare$\medskip

\begin{lemma}
	\label{lemm: unif int}
	Suppose that $n \ge 2$, and $\Sigma_n$ is a block diagonal matrix along a cluster structure $\mathcal{M}_n$, where the $m$-th block, denoted by $\Sigma_{n,m}$ is given by
	\begin{align*}
		\Sigma_{n,m} = \sigma^2 (1 -\delta_{n,m})I_{n_m} + \sigma^2 \delta_{n,m} \mathbf{1}_{n_m} \mathbf{1}_{n_m}^\top,
	\end{align*}
	where
	\begin{eqnarray}
		\label{bound312}
		\delta_{n,m} = \frac{\overline \delta}{n^*},
	\end{eqnarray}
	for some $\overline\delta \in [-a,a]$, with $0<a<1/2$, if $n_m \ge 2$, and $\sigma^2>0$ is independent of $n$.

	Then, the following holds for any random vector $X_n \in \mathbf{R}^n$ which follows $\Phi(0,\sigma^2 I_n)$.

	(i) $(d\Phi(0,\Sigma_n)/d\Phi(0,\sigma^2 I_n))(X_n)$ is uniformly integrable.

	(ii) $\log\left( (d\Phi(0,\Sigma_n)/d\Phi(0,\sigma^2 I_n))(X_n) \right)$ is uniformly tight.
\end{lemma}

\noindent \textbf{Proof: } For brevity, we focus on the case $\sigma^2 = 1$. By Corollary \ref{cor: log LR cluster},
\begin{align*}
	\log \frac{d \Phi(0,\Sigma_n)}{d \Phi(0,I_n)}(X_n) = A_n + R_n,
\end{align*}
where
\begin{align*}
	A_n&= - \sum_{m=1}^{M_n} \log \sqrt{1 + (n_m-1) \delta_{n,m}} + \sum_{m=1}^{M_n} \frac{(n_m-1) \delta_{n,m} Z_{m,i_m}^2}{2(1 + (n_m-1) \delta_{n,m})}, \text{ and }\\
	R_n&= - \sum_{m=1}^{M_n} (n_m-1) \log \sqrt{1 - \delta_{n,m}}- \sum_{m=1}^{M_n} \frac{\delta_{n,m}}{2(1 - \delta_{n,m})} \sum_{i \in N_{n,m} \setminus \{i_m\}} Z_{m,i}^2,
\end{align*}
and $i_m$ denotes the first $i$ in block $m$.\medskip

\noindent (i) Let us take small $\epsilon>0$ such that
\begin{align}
	\label{bound}
	a < \frac{2}{(1 + \epsilon)(4 + \epsilon)}.
\end{align}
We write (under $\Phi(0,I_n)$)
\begin{align*}
	\mathbf{E}\left[ \left(\frac{d \Phi(0,\Sigma_n)}{d \Phi(0,I_n)} \right)^{(1 + \epsilon)}\right] = \mathbf{E}\left[\exp((1 + \epsilon)A_n) \right]\mathbf{E}\left[\exp(( 1 + \epsilon)R_n) \right],
\end{align*}
since $A_n$ and $R_n$ are independent.  Let $t_m = (n_m-1)/n^*$, and write
\begin{align*}
	\mathbf{E}\left[\exp((1 + \epsilon)A_n) \right] &= \prod_{m=1}^{M_n} \left( 1 + t_m\overline \delta\right)^{-\frac{1 + \epsilon}{2}}
	\mathbf{E}\left[ \exp\left(\frac{t_m  (1 + \epsilon) \overline \delta}{2(1 + t_m \overline \delta)} Z_{m,1}^2\right)\right]\\
	&\le \prod_{m=1}^{M_n} \left( 1 + t_m\overline \delta\right)^{-\frac{1 + \epsilon}{2}} \left(1 - t_m (1 + \epsilon) \overline \delta \right)^{-1/2} \equiv f_n(\overline \delta), \text{ say}.
\end{align*}
Note that
\begin{align*}
	\frac{d \log f_n(\overline \delta)}{d \overline \delta} = \frac{(1 + \epsilon)(2 + \epsilon) \overline \delta}{2}\sum_{m=1}^{M_n}\frac{t_m^2}{(1 + t_m \overline \delta)( 1 - t_m \overline \delta ( 1 + \epsilon))} >0,
\end{align*}
because $t_m \le 1$ and $\overline \delta ( 1+ \epsilon) <1$ by (\ref{bound}). This means that $f_n(\overline \delta)$ is increasing in $\overline \delta \in [-a,a]$ and achieves its maximum at $\overline \delta = a$. Hence,
\begin{align*}
	\left| \frac{d \log f_n(\overline \delta)}{d \overline \delta}\right| a \le \left| \frac{d \log f_n(\overline \delta)}{d \overline \delta}\right| &\le \frac{(1 + \epsilon)(2 + \epsilon) \overline \delta}{2( 1- (1+ \epsilon) \overline \delta)}\sum_{m=1}^{M_n} t_m^2 \\
	&\le \frac{(1 + \epsilon)(2 + \epsilon) a}{2( 1- (1+ \epsilon) a)} < 1,
\end{align*}
because $\sum_{m=1}^{M_n} t_m^2 \le 1$ and we chose $\epsilon$ such that (\ref{bound}) holds. By Lemma \ref{lemm: bound basic}, we have
\begin{align*}
	f_n(\overline \delta) \le f_n(a) \le \left( 1 - \frac{(1 + \epsilon)(2 + \epsilon) a}{2( 1- (1+ \epsilon) a)} \right)^{-1}.
\end{align*}
The bound does not depend on $n$, and hence,
\begin{align*}
	\sup_{n \ge 1} \mathbf{E}\left[\exp((1 + \epsilon)A_n) \right] < \infty.
\end{align*}

Now, we turn to $ \mathbf{E}\left[\exp((1 + \epsilon)R_n) \right]$. We can write
\begin{align}
	\label{Rn}
	R_n = -\frac{1}{2} n^* \log\left( 1 - \frac{\overline \delta}{n^*} \right) - \frac{1}{2} \frac{\overline \delta /n^*}{1 - \overline \delta/n^*} \sum_{m=1: n_m \ne 1}^{M_n} \sum_{i \in N_{n,m} \setminus\{i_m\}} Z_{m,i}^2.
\end{align}
Using this expression, we rewrite
\begin{align*}
	\mathbf{E}\left[ \exp\left( (1 + \epsilon) R_n \right)\right] &= \left( 1 - \frac{\overline \delta}{n^*} \right)^{-\frac{n^*(1 + \epsilon)}{2}} \prod_{m=1: n_m \ne 1}^{M_n} \prod_{i \in N_{n,m} \setminus \{i_m\}} \mathbf{E}\left[ \exp\left( -\frac{1}{2} \frac{\overline \delta ( 1+ \epsilon)/n^*}{1 - \overline \delta/n^*} Z_{m,i}^2\right)\right]\\
	& = \left( 1 - \frac{\overline \delta}{n^*} \right)^{-\frac{n^*(1 + \epsilon)}{2}} \prod_{m=1: n_m \ne 1}^{M_n} \prod_{i \in N_{n,m} \setminus \{i_m\}} \left( 1 + \frac{\overline \delta ( 1+ \epsilon)/n^*}{1 - \overline \delta/n^*} \right)^{-\frac{1}{2}} \\
	&\le \left( 1 - \frac{\overline \delta}{n^*} \right)^{-\frac{n^*(1 + \epsilon)}{2}} \le \left( 1 - \frac{a}{n^*} \right)^{-\frac{n^*(1 + \epsilon)}{2}}.
\end{align*}
The last bound is a sequence converging to $\exp(a(1 + \epsilon)/2)$ as $n^* \rightarrow \infty$, and Hence, is a bounded sequence. Thus, we conclude that
\begin{align*}
	\sup_{n \ge 1} \mathbf{E}\left[\exp((1 + \epsilon)R_n) \right] < \infty.
\end{align*}
This proves that
\begin{align*}
	\sup_{n \ge 1} \mathbf{E}\left[ \left(\frac{d \Phi(0,\Sigma_n)}{d \Phi(0,I_n)} \right)^{(1 + \epsilon)}(X_n)\right] < \infty.
\end{align*}
Hence, the proof of (i) is complete.\medskip

\noindent (ii) We rewrite
\begin{align}
	\label{sum}
	A_n = - \frac{1}{2}\sum_{m=1}^{M_n}\left(\log\left( 1 + (n_m - 1) \delta_{n,m}\right) - \frac{(n_m-1)\delta_{n,m}}{1 + (n_m-1)\delta_{n,m}} \right)
	+ \sum_{m=1}^{M_n} \frac{(n_m-1) \delta_{n,m}(Z_{m,i_m}^2 - 1)}{2(1 + (n_m-1) \delta_{n,m})}.
\end{align}
For any $x \in [0,1]$, we have
\begin{eqnarray*}
	\frac{x^2}{8} \le	\frac{1}{2} \frac{x^2}{(1 + x)^2} \le \log(1+x) - \frac{x}{1+x} \le \frac{1}{2} \frac{x^2}{1 + x} \le \frac{x^2}{2}.
\end{eqnarray*}
Hence,
\begin{align*}
	\left| \frac{1}{2}\sum_{m=1}^{M_n}\left(\log\left( 1 + (n_m - 1) \delta_{n,m}\right) - \frac{(n_m-1)\delta_{n,m}}{1 + (n_m-1)\delta_{n,m}} \right)\right|
	\le \frac{\overline \delta^2}{2}\sum_{m=1}^{M_n} \frac{(n_m-1)^2 }{2n^{*2}} \le \frac{\overline \delta^2}{2}.
\end{align*}
It suffices to show the uniform tightness of the second sum in (\ref{sum}). Under $\Phi(0,I_n)$, it has mean zero, and
\begin{align*}
	\Var\left( \sum_{m=1: n_m \ge 2}^{M_n}  \frac{\overline \delta (n_m-1)/n^*}{1 + (\overline \delta (n_m-1)/n^*)}(Z_{m,i_m}^2 - 1)\right) &\le \sum_{m=1}^{M_n} \frac{(\overline \delta (n_m-1)/n^*)^2}{(1 + (\overline \delta (n_m-1)/n^*))^2} \Var\left( Z_{m,i_m}^2 \right) \\
	&\le 2 \overline \delta^2 \sum_{m=1: n_m \ge 2}^{M_n}\left(\frac{n_m-1}{n^*}\right)^2 \le 2 \overline \delta^2,
\end{align*}
because $\Var(Z_{m,i_m}^2) = 2$. Therefore, $A_n$ is uniformly tight.

As for $R_n$, we recall (\ref{Rn}), and can follow similar arguments to show that $R_n$ is uniformly tight as well. $\blacksquare$

\subsection{Consistent Discrimination of the Mean}

We let
\begin{align}
	\label{Sigma delta}
	\Sigma(\sigma^2,\delta) = \sigma^2\left(I_n + \delta \mathbf{1}_n \mathbf{1}_n^\top - \delta I_n\right),
\end{align}
and for any $\overline \mu \in \mathbf{R}$, we write $\Phi(\overline \mu \mathbf{1}_n,\Sigma(\sigma^2,\delta))$ simply as $\Phi(\overline \mu,\sigma^2,\delta)$. Let us recall some basic notions of optimality of tests \citep{Lehmann/Romano:05:TSH}. Given a model $\mathcal{P}_{n}$ which is partitioned as $\mathcal{P}_{n,0} \cup \mathcal{P}_{n,1}$, a test $\phi_n$ is said to be a \bi{UMP (uniformly most powerful) test} of $\mathcal{P}_{n,0}$ against $\mathcal{P}_{n,1}$ at level $\alpha \in (0,1)$, if under any $P_{n,0} \in \mathcal{P}_{n,0}$,
\begin{align*}
	\mathbf{E}[\phi_n] \le \alpha,
\end{align*}
and for any alternative test $\phi_n'$ such that $\mathbf{E}[\phi_n'] \le \alpha$ under any $P_{n,0} \in \mathcal{P}_{n,0}$, we have
\begin{align*}
	\mathbf{E}[\phi_n'] - \mathbf{E}[\phi_n] \le 0,
\end{align*}
under any $P_{n,1} \in \mathcal{P}_{n,1}$.

A sequence of tests $\phi_n$ is said to be an \bi{AUMP (asymptotically uniformly most powerful) test} of $\mathcal{P}_{n,0}$ against $\mathcal{P}_{n,1}$ at level $\alpha \in (0,1)$, if under any sequence $P_{n,0} \in \mathcal{P}_{n,0}$,
\begin{align*}
	\limsup_{n \rightarrow \infty} \mathbf{E}[\phi_n] \le \alpha,
\end{align*}
and for any alternative test $\phi_n'$ such that $\limsup_{n \rightarrow \infty} \mathbf{E}[\phi_n'] \le \alpha$ under any sequence $P_{n,0} \in \mathcal{P}_{n,0}$, we have
\begin{align*}
	\limsup_{n \rightarrow \infty} \mathbf{E}[\phi_n'] - \mathbf{E}[\phi_n] \le 0,
\end{align*}
under any sequence $P_{n,1} \in \mathcal{P}_{n,1}$.

\begin{lemma}
	\label{lemm: UMP AUMP}
	Suppose that $A \subset \mathbf{R}$ is an open interval, and $\{\varphi_\alpha\}_{\alpha \in A}$ is a class of tests of $\mathcal{P}_{n,0}$ against $\mathcal{P}_{n,1}$, such that for each $\alpha \in A$, the test $\varphi_\alpha$ is UMP at level $\alpha$, and for any $\tilde \alpha = \alpha +o(1)$ as $n \rightarrow \infty$,
	\begin{align}
		\label{cond}
		\mathbf{E}[\varphi_{\alpha}] = \mathbf{E}[\varphi_{\tilde \alpha}] + o(1),
	\end{align}
	under any sequence $P_n \in \mathcal{P}_{n,0} \cup \mathcal{P}_{n,1}$. Then, $\varphi_\alpha$ is AUMP at level $\alpha$.
\end{lemma}

\noindent \textbf{Proof: } Choose any test $\tilde \varphi$ such that under any sequence $P_n \in \mathcal{P}_{n,0}$,
\begin{align*}
	\mathbf{E}[\tilde \varphi] \le \alpha + \epsilon_n,
\end{align*}
for some sequence $\epsilon_n \rightarrow 0$, as $n \rightarrow \infty$. Fix one such sequence $P_n\in \mathcal{P}_{n,0}$, together with the sequence $\epsilon_n$, and let $\tilde \alpha = \alpha + \epsilon_n$. Now, select a large enough $n$ such that $\tilde \alpha \in A$ and choose any $P_n' \in \mathcal{P}_{n,1}$. Then, since $\varphi_{\tilde \alpha}$ is UMP at level $\tilde \alpha$, we have
\begin{align*}
	\mathbf{E}_{P_n'}[\varphi_{\tilde \alpha}] \ge \mathbf{E}_{P_n'}[\tilde \varphi]
\end{align*}
under $P_n'$. By (\ref{cond}), we can see that $\varphi_{\alpha}$ is AUMP at level $\alpha$. $\blacksquare$\medskip

\noindent \textbf{Proof of Theorem \ref{thm: consistent discrimination}: } Suppose that $M_n = 1$. First, consider the case where $\mathcal{P}_{n} = \mathcal{P}_{n,\mathcal{N}}'$, with
\begin{align*}
	\mathcal{P}_{n,\mathcal{N}}' = \left\{\Phi(\mu,\sigma^2,\delta): (n-1)\delta \in (0,1), \sigma^2>0, \mu \ge 0\right\}.
\end{align*}
Later, we generalize the result to the case where $\mathcal{P}_n$ contains the above probability model. Define $\mathcal{P}_{n,0} = \left\{\Phi(0,\sigma^2,\delta): (n-1) \delta \in (0,1), \sigma^2>0 \right\}$ and let $\mathcal{P}_{n,1} = \mathcal{P}_{n}  \setminus \mathcal{P}_{n,0}$.
In light of Lemma \ref{lemm: UMP AUMP}, it suffices to construct a class of tests $\{\varphi_{\alpha}\}_{\alpha \in (0,1/2)}$ of $\mathcal{P}_{n,0}$ against $\mathcal{P}_{n,1}$ such that

(a) it satisfies (\ref{cond}) in Lemma \ref{lemm: UMP AUMP},

(b) the test $\varphi_{\alpha}$ has power bounded by a constant below $1$ uniformly over $n$, and

(c) each test $\varphi_{\alpha}$ is a UMP test of $\mathcal{P}_{n,0}$ against $\mathcal{P}_{n,1}$.

Let us first construct such a test and show that (a)-(c) are satisfied. Define
\begin{align*}
	V_n = \frac{Z_{n,1}}{|Z_{n,1}|} = \text{sgn}(Z_{n,1}).
\end{align*}
For each $\alpha \in (0,1/2)$, let
\begin{align*}
	\phi_\alpha(V_n) = \left\{\begin{array}{ll}
		1,& \text{ if } V_n > C_0\\
		\gamma_0, & \text{ if } V_n = C_0,\\
		0, & \text{ if } V_n < C_0,
	\end{array}
	\right.
\end{align*}
for some $C_0>0$ and $\gamma_0 \in [0,1]$. Let $Z = (Z_{n,1} - \mathbf{E}[Z_{n,1}])/\sqrt{\Var(Z_{n,1})}$. Then the size control requires that under the null hypothesis,
\begin{align}
	\label{size control2}
	\mathbf{E}[\phi_\alpha(V_n)] = P\{Z > C_0|Z|\} + P\{Z = C_0|Z|\} \gamma_0 = \alpha.
\end{align}
Since $\alpha \in (0,1/2)$, we must have $C_0 = 1$ and $\gamma_0 = 2 \alpha$.

Let us first show that this test satisfies the condition (a). For any $\tilde \alpha$ such that $\tilde \alpha = \alpha + o(1)$, and under any sequence $P_n \in \mathcal{P}_{n}$,
\begin{align}
	\label{size control3}
	\mathbf{E}[\phi_{\tilde \alpha}(V_n)] = 2 \tilde \alpha P_n\{Z_{n,1} \ge 0\} = 2 \alpha P_n\{Z_{n,1} \ge 0\} + o(1) = \mathbf{E}[\phi_{\alpha}(V_n)] + o(1).
\end{align}
Hence, the class of tests $\{\phi_{\alpha}(V_n)\}_{\alpha \in (0,1/2)}$ satisfies the condition (\ref{cond}).

As for the condition (b), note that under any alternative hypothesis in $\mathcal{P}_{n,1}$, we have
\begin{align*}
	\mathbf{E}[\varphi_\alpha(V_n)] = P_n\{Z_{n,1} > |Z_{n,1}|\} + 2 \alpha P_n\{Z_{n,1} = |Z_{n,1}|\} = 2 \alpha P_n\{Z_{n,1} \ge 0 \} \le 2\alpha.
\end{align*}
Hence, the test does not have power exceeding $2\alpha < 1$.

Finally, we show that the condition (c) is satisfied. Let
\begin{align}
	\label{LX}
	\mathcal{L}(X_n;\mu,\sigma^2,\delta_n) &= - \log \sqrt{\sigma^2(1 + (n-1) \delta_n)} - (n-1) \log \sqrt{\sigma^2 (1 - \delta_n)} \\ \notag
	&\quad \quad  +\frac{\sigma^2 (1 + (n-1) \delta_n)-1}{2 \sigma^2 (1 + (n-1) \delta_n)} Z_{n,1}^2 +\frac{\sigma^2 (1-\delta_n)-1}{2 \sigma^2 (1 - \delta_n)} \sum_{k=2}^n Z_{n,k}^2 \\ \notag
	&\quad \quad  + \sqrt{n} \mu Z_{n,1} - \frac{n \mu^2 \sigma^2 (1 + (n-1)\delta_n)}{2}.
\end{align}
Hence, $\mathcal{L}(X_n;\mu,\sigma^2,\delta_n)$ is the same as $\log\left(d \Phi(\mu \mathbf{1}_n,\Sigma(\sigma^2,\delta_n)/d \Phi(0,I_n)\right)(X_n)$ in Lemma \ref{lemma: LR for the mean}, except that the coefficient of $Z_{n,1}$ is $\sqrt{n}\mu$ and the last term is different. Define a probability measure $P_n(\mu,\delta_n)$ as follows: for any Borel $B$,
\begin{align*}
	P_n(\mu,\delta_n)(B) = \int_B \exp\left(\mathcal{L}(x; \mu,\delta_n)\right)d\Phi(0,I_n)(x).
\end{align*}
Similarly as before, we define
\begin{align*}
	\mathcal{\tilde P}_{n} = \left\{P_n(\mu,\delta): (n-1)\delta \in (0,1), \mu \ge 0\right\}, \mathcal{\tilde P}_{n,0} = \left\{P_n(0,\delta): (n-1)\delta \in (0,1) \right\},
\end{align*}
and let $\mathcal{\tilde P}_{n,1} = \mathcal{\tilde P}_{n}  \setminus \mathcal{\tilde P}_{n,0}$. It is not hard to see that
\begin{align*}
	\mathcal{\tilde P}_{n,0} = \mathcal{P}_{n,0} \quad \text{ and } \quad \mathcal{\tilde P}_{n,1} = \mathcal{P}_{n,1}.
\end{align*}
Therefore, a UMP test of $\mathcal{\tilde P}_{n,0}$ against $\mathcal{\tilde P}_{n,1}$ is also a UMP test of $\mathcal{P}_{n,0}$ against $\mathcal{P}_{n,1}$. It suffices for condition (c) to show that the test $\varphi_\alpha$ is a UMP test of $\mathcal{\tilde P}_{n,0}$ against $\mathcal{\tilde P}_{n,1}$. From (\ref{LX}), the sufficient statistics for $\mathcal{\tilde P}_{n}$ in the case of $M_n = 1$ are given by
\begin{align*}
	\left(Z_{n,1}, Z_{n,1}^2, \sum_{k=2}^n Z_{n,k}^2\right),
\end{align*}
where $Z_{n,k}$'s are as in Lemma \ref{lemma: LR for the mean}. For any $t \ge 0$,
\begin{align*}
	P\left\{V_n = 1 , Z_{n,1}^2 \le t \right\} &= P\left\{V_n = 1 , -\sqrt{t} \le Z_{n,1} \le \sqrt{t} \right\}
	= P\left\{Z_{n,1} > 0, -\sqrt{t} \le Z_{n,1} \le \sqrt{t} \right\} \\
	&= P \left\{ 0 \le Z_{n,1} \le \sqrt{t} \right\} = \frac{1}{2}P\left\{- \sqrt{t} \le Z_{n,1} \le \sqrt{t}\right\} = P\left\{V_n = 1\right\}P\left\{Z_{n,1}^2 \le t \right\},
\end{align*}
under the null hypothesis. Hence, $V_n$ and $Z_{n,1}^2$ are independent under any probability in $\mathcal{\tilde P}_{n,0}$. Furthermore, under any probability in $\mathcal{\tilde P}_{n,0}$,
\begin{align*}
	\mathbf{E}\left[Z_n Z_n^\top\right] &= B_n^\top \mathbf{E}\left[Z_n Z_n^\top\right] B_n\\
	&=B_n^\top \Sigma_n B_n = B_n^\top (I_n + \delta_n \mathbf{1}_n \mathbf{1}_n^\top - \delta_n I_n) B_n = (1 - \delta_n)I_n + B_n^\top \mathbf{1}_n \mathbf{1}_n B_n.
\end{align*}
Note that $B_n^\top \mathbf{1}_n \mathbf{1}_n B_n$ is a matrix whose $(1,1)$-th entry is $b_1^\top  \mathbf{1}_n \mathbf{1}_n^\top b_1 = 1$ and all the other entries are zeros. Hence, $Z_{n,k}$'s are independent across $k$'s under any probability in $\mathcal{\tilde P}_{n,0}$. Therefore, $V_n$ and $(Z_{n,1}^2,\sum_{k=2}^n Z_{n,k}^2)$ are independent under any probability in $\mathcal{\tilde P}_{n,0}$. By Theorem 5.1.1 of \cite{Lehmann/Romano:05:TSH}, the randomized test $\varphi_\alpha(V_n)$ is an $\alpha$-level UMP test. 

Next, consider the case where $\mathcal{P}_{n,\mathcal{N}}' \subset \mathcal{P}_n$. Take a sequence of tests $\varphi_n$ such that for any sequence of probabilities $P_n \in \mathcal{P}_{n,0}$, $\limsup_{n \rightarrow \infty} \mathbf{E}[\varphi_n(X_n)] \le \alpha$. Now, we take a sequence $P_{n,1} \in \mathcal{P}_{n,1} \cap \mathcal{P}_{n,\mathcal{N}}'$. For any $\alpha \in (0,1/2)$, the test $\phi_\alpha(V_n)$ is a UMP test at level $\alpha$ of the null hypothesis $\mathcal{P}_{n,0} \cap \mathcal{P}_{n,\mathcal{N}}'$ against $\mathcal{P}_{n,1} \cap \mathcal{P}_{n,\mathcal{N}}'$. Note that for any sequence of probabilities $P_n \in \mathcal{P}_{n,0} \cap \mathcal{P}_{n,\mathcal{N}}'$, $\limsup_{n \rightarrow \infty} \mathbf{E}[\varphi_n(X_n)] \le \alpha$. Hence, if we take $\epsilon>0$ such that $\alpha + \epsilon \in (0,1/2)$, there exists $n_0 \ge 1$ such that for all $n \ge n_0$, $\mathbf{E}[\varphi_n(X_n)] \le \alpha + \epsilon$. For all such $n$, under any $P_n \in \mathcal{P}_{n,1} \cap \mathcal{P}_{n,\mathcal{N}}'$, we have
\begin{align*}
	\mathbf{E}[\varphi_n(X_n)] \le \mathbf{E}[\phi_{\alpha + \epsilon}(V_n)] \le 2 (\alpha + \epsilon).  
\end{align*}
Hence, we find that along any sequence $P_n \in \mathcal{P}_{n,1} \cap \mathcal{P}_{n,\mathcal{N}}'$, we have 
\begin{align*}
	\liminf_{n \rightarrow \infty }\mathbf{E}[\varphi_n(X_n)] \le 2 (\alpha + \epsilon) < 1.
\end{align*}
Thus, the proof is complete. $\blacksquare$\medskip

The following lemma is used for the proof of Theorem \ref{thm: root n discr}. For $w \in [0,1]$, define $\xi_1 = w Z_1$ and $\xi_2 = (1-w)Z_2$, where $Z_i \sim N(0,1)$, independent across $i=1,2$. Define
\begin{align*}
	T(w) = \frac{(\xi_1 + \xi_2)/\sqrt{2}}{\sqrt{(\xi_1 - \bar \xi)^2 + (\xi_2 - \bar \xi)^2}},
\end{align*}
where $\bar \xi  = (\xi_1 + \xi_2)/2$. Let us take $t \ge 0$, and define
\begin{align*}
	p(w;t) = P\{T(w) \le t\}.
\end{align*}
\begin{lemma}
	\label{lemm: t distr}
	(i) For all $t \ge 1$ and all $w \in [0,1]$, $p(w;t) \ge p(1/2;t)$.

	(ii) For all $0\le t<1$ and all $w \in [0,1]$, $p(w;t) \le p(1/2;t)$.
\end{lemma}

\noindent \textbf{Proof: } First, we write
\begin{align*}
	T(w) = \frac{\xi_1 + \xi_2}{\sqrt{(\xi_1 - \xi_2)^2}}.
\end{align*}
For (i) and (ii), since $[\xi_1,\xi_2]$ is symmetrically distributed around the origin, it suffices to show that $p(\,\cdot\,;t)$ is increasing on $[1/2,1]$ for all $t \ge 1$, and $p(\,\cdot\,;t)$ is decreasing on $[1/2,1]$ for all $0 \le t<1$. Let $A$ denote the event that $\xi_1 + \xi_2 <0$. Then, on the event $A$, for all $w \ge 0$ and all $t \ge 0$, we have $T(w) \le t$. Hence,
\begin{align*}
	P\left(\{T(w) \le t\} \cap A\right) = P(A) = 0.5.
\end{align*}
On the event $A^c$, $T(w) \le t$ if and only if
\begin{align*}
	w^2 Z_1^2 + (1-w)^2 Z_2^2 + 2 w(1-w) Z_1 Z_2 \le (w^2 Z_1^2 + (1-w)^2 Z_2^2 - 2 w(1-w)Z_1 Z_2)t^2
\end{align*}
if and only if
\begin{align*}
	0 \le (w^2 Z_1^2 + (1-w)^2 Z_2^2)(t^2 -1) - 2(t^2 +1) w(1-w)Z_1 Z_2 = f(w;Z_1,Z_2),
\end{align*}
where
\begin{align*}
	f(w;Z_1,Z_2) = (w Z_1 + (1-w)Z_2)^2(t^2 - 1) - 4 t^2 w(1-w) Z_1 Z_2.
\end{align*}

Take $t \ge 1$. Let $B$ be the event $Z_1 Z_2 \le 0$. Certainly, if $Z_1 Z_2 \le 0$, $f(w; Z_1,Z_2) \ge 0$ for all $w \in [1/2,1]$. Hence
\begin{align*}
	P\left(\{T(w) \le t\} \cap A^c \right)&= P\left(\{T(w) \le t\} \cap A^c \cap B\right) + P\left(\{T(w) \le t\} \cap A^c \cap B^c\right) \\
	&=  P\left(A^c \cap B \right) + P\left(\{T(w) \le t\} \cap A^c \cap B^c \right).
\end{align*}
We show that $f(w;Z_1,Z_2)$ is increasing in $w$ on the event $A^c$.  We take the derivative $f'(w;Z_1,Z_2)$ with respect to $w$:
\begin{align*}
	f'(w;Z_1,Z_2) = 2(wZ_1 - (1-w) Z_2)(t^2-1) (Z_1 - Z_2) - 4 t^2 (1-2w) Z_1Z_2.
\end{align*}
The function is linear in $w$. First, we take $w =1$. Then, on the event $B^c$,
\begin{align*}
	f'(1;Z_1,Z_2) &= 2(t^2-1)Z_1 (Z_1 - Z_2) + 4t^2 Z_1 Z_2 = 2 (t^2-1) Z_1^2 + (4t^2 - (2(t^2-1))Z_1 Z_2\\
	  &=2 (t^2-1) Z_1^2 + (2t^2 + 2)Z_1 Z_2 \ge 0.
\end{align*}
Second, we take $w=1/2$. Then,
\begin{align*}
	f'(1/2;Z_1,Z_2) = (t^2-1) (Z_1 - Z_2)^2 .
\end{align*}
Hence, for all $Z_1,Z_2$ such that $Z_1 Z_2 > 0$, $f(w;Z_1,Z_2)$ is increasing on $[1/2,1]$ for all $t \ge 1$. Therefore, whenever $t \ge 1$, $P\{T(w) \le t\}$ is increasing on $[1/2,1]$.

Take $0 \le t <1$.  If $Z_1 Z_2 > 0$, then $f(w;Z_1,Z_2) < 0$. Hence
\begin{align*}
	P\left(\{T(w) \le t\} \cap A \right)&= P\left(\{T(w) \le t\} \cap A^c \cap B \right) + P\left(\{T(w) \le t\} \cap A^c \cap B^c \right) \\
	&=  P\left(\{T(w) \le t\} \cap A^c \cap B \right).
\end{align*}
On the event $B$, when $w=1$,
\begin{align*}
	f'(1;Z_1,Z_2) = 2 (t^2-1) Z_1^2 + (2t^2 + 2)Z_1 Z_2 \le 0,
\end{align*}
and when $w=1/2$,
\begin{align}
	f'(1/2;Z_1,Z_2) = (t^2-1) (Z_1 - Z_2)^2\le 0.
\end{align}
Hence, for all $Z_1,Z_2$ such that $Z_1 Z_2 \le 0$, $f(w;Z_1,Z_2)$ is decreasing on $[1/2,1]$ for all $0 \le t <1$. Therefore, whenever $0 \le t < 1$, $P\{T(w) \le t\}$ is decreasing on $[1/2,1]$. $\blacksquare$\medskip

\noindent \textbf{Proof of Theorem \ref{thm: root n discr}: } Suppose that we have at least two nonnegligible clusters, i.e., $M_n \ge 2$ for all but finite number of $n$'s. Consider testing the null hypothesis of $\mathbf{E}[X_{n,i}] = 0$ against $\mathbf{E}[X_{n,i}] > 0$. Without loss of generality, we enumerate $\mathcal{M}_n' = \{N_1,...,N_{M_n'}\}$, $M_n' \ge 2$. Now, we construct a test that consistently $\sqrt{n}$-discriminates the mean. Define
\begin{align*}
	\xi_m = \frac{1}{\sqrt{n_m}}\sum_{i \in N_{n,m}} X_{n,i},
\end{align*}
and
\begin{align*}
	U_n' = \frac{1}{2} \sum_{m=1}^{2} \xi_m  \quad\text{and}\quad T_n' = \sum_{m=1}^{2} \left(\xi_m - \frac{1}{2} \sum_{m=1}^{2} \xi_m \right)^2.
\end{align*}
We take
\begin{align*}
	V_n' = \frac{\sqrt{2} U_n'}{\sqrt{T_n'}}.
\end{align*}
Let $c_{1 - \alpha}$ be the $1 - \alpha$ quantile of the $t$-distribution with degree of freedom $1$. Define
\begin{align*}
	\varphi_n = 1\{V_n' > \max\{c_{0,75},c_{1 - \alpha}\}\}.
\end{align*}
Note that $c_{0.75} = 1$.

We first show that this test controls the size of the test under $\alpha$ asymptotically under the null hypothesis. Define an infeasible test statistic
\begin{align*}
	V_n'' = \frac{\sqrt{2} U_n''}{\sqrt{T_n''}},
\end{align*}
where
\begin{align*}
	U_n'' = \frac{1}{2} \sum_{m=1}^{2}\frac{\xi_m}{\sigma_{n,m}} \quad\text{and}\quad
	T_n'' = \sum_{m=1}^{2} \left(\frac{\xi_m}{\sigma_{n,m}} - \frac{1}{2} \sum_{m=1}^{2} \frac{\xi_m}{\sigma_{n,m}} \right)^2.
\end{align*}
Here $\sigma_{n,m}^2$ is the variance of $\xi_m$ under the null hypothesis. Then $V_n''$ converges in distribution to the $t$-distribution with $1$ degree of freedom under the null hypothesis. By Lemma \ref{lemm: t distr}, if $0< \alpha < 0.25$ so that $c_{1-\alpha} >1$,
\begin{align}
	\mathbf{E} [\varphi_n] = P\left\{ V_n' > c_{1 - \alpha}\right\} \le P\left\{ V_n'' > c_{1 -\alpha}\right\} = \alpha + o(1),
\end{align}
and if $\alpha \ge 0.25$ so that $0 \le c_{1-\alpha} \le 1$,
\begin{align}
	\mathbf{E} [\varphi_n] = 1 -  P\left\{ V_n' \le c_{0.75}\right\} \le 1 - P\left\{ V_n'' \le c_{0.75}\right\} = 0.25 \le \alpha + o(1),
\end{align}
as $n \rightarrow \infty$. Hence, the size of the test $\varphi_n$ is bounded by $\alpha$ asymptotically.

Suppose that we are under the local alternatives such that $\mathbf{E}[X_n] = \overline \mu_n \sigma_{LR}\mathbf{1}_n/\sqrt{n}$, for some sequence $\overline \mu_n \rightarrow \infty$. Define
\begin{align*}
	\bar U_n = U_n' - \frac{1}{2}\sum_{m=1}^{2} \sqrt{\frac{n_m}{n}} \overline \mu_n \sigma_{LR}.
\end{align*}
Note that
\begin{align}
	\label{last prob}
	P\{V_n' > c_{1-\alpha}\} &= P\left\{\frac{\sqrt{2} \bar U_n}{\sigma_{LR}} > c_{1-\alpha} \sqrt{\frac{T_n'}{\sigma_{LR}^2}} -  \frac{1}{\sqrt{2}}\sum_{m=1}^{2} \sqrt{\frac{n_m}{n}} \overline \mu_n \right\}\\ \notag
	&\ge P\left\{\frac{\sqrt{2} \bar U_n}{\sigma_{LR}} > c_{1-\alpha} \sqrt{\frac{T_n'}{\sigma_{LR}^2}} -  \frac{1}{\sqrt{2}}\sum_{m=1}^{2} \frac{n_m}{n} \overline \mu_n  \right\},
\end{align}
because $n_m /n \le 1$. Note that
\begin{align}
	\sigma_{LR}^2 = \sum_{m =1}^{M_n} \frac{n_m}{n}\sigma_{n,m}^2 \ge \sum_{m =1}^{2} \frac{n_m}{n}\sigma_{n,m}^2.
\end{align}
Since
\begin{align*}
	0 < \liminf_{n \rightarrow \infty} \min_{m=1,2} \frac{n_m}{n} \le \limsup_{n \rightarrow \infty} \min_{m=1,2} \frac{n_m}{n} \le 1,
\end{align*}
there exist $\epsilon>0$ and $n_0>0$ such that for all $n \ge n_0$,
\begin{align*}
	\sigma_{LR}^2 \ge \epsilon \sum_{m =1}^{2} \sigma_{n,m}^2 \ge \epsilon \max_{m=1,2} \sigma_{n,m}^2.
\end{align*}
Therefore,
\begin{align*}
	\frac{|\bar U_n|}{\sigma_{LR}} \le \frac{\max_{m=1,2}\sigma_{n,m}}{2 \sigma_{LR}} \sum_{m=1}^{2} \frac{\left| \xi_m - \mathbf{E}[\xi_m]\right|}{\sigma_{n,m}} \le \frac{1}{2 \sqrt{\epsilon}} \sum_{m=1}^{2} \frac{\left| \xi_m - \mathbf{E}[\xi_m]\right|}{\sigma_{n,m}}.
\end{align*}
Since $(\xi_m - \mathbf{E}\xi_m)/\sigma_{n,m}$ converges in distribution to $N(0,1)$ under any sequence $P_n \in \mathcal{P}_n$ as $n \rightarrow \infty$ by (\ref{conv2}), we have $\bar U_n /\sigma_{LR} = O_P(1)$. Similarly, we can show that  $T_n'/\sigma_{LR}^2 = O_P(1)$. Hence, the last probability in (\ref{last prob}) converges to one as $\overline \mu_n \rightarrow \infty$, proving that the mean is consistently $\sqrt{n}$-discriminated.  $\blacksquare$

\subsection{Impossibility of Consistent Estimation of Long Run Variance}

\noindent \textbf{Proof of Lemma \ref{lemm: basic}: } We first show sufficiency. Suppose that $\|\theta_{n}(P_{n,1}) - \theta_{n}(P_{n,0})\| = o(1)$ for any sequence $P_{n,1} \in \mathcal{P}_n$. Then we take $\hat \theta = \theta_n(P_{n,0})$, so that $\| \hat \theta - \theta_n(P_{n,1})\| \rightarrow 0$, along $P_{n,1} \in \mathcal{P}_n$, as $n \rightarrow \infty$. Hence, sufficiency follows.

Conversely, suppose that $\theta_n$ is consistently estimable in $\mathcal{P}_n$, so that there exists an estimator, say, $\tilde \theta$, such that $\tilde \theta - \theta_n(P_{n,1}) = o_P(1)$ along any $P_{n,1} \in \mathcal{P}_n$. Since $P_{n,0} \in \mathcal{P}_n$, this means that $\tilde \theta - \theta_n(P_{n,0}) = o_P(1)$ under $P_{n,0}$. Since $P_{n,1} \triangleleft P_{n,0}$, $\tilde \theta - \theta_n(P_{n,0}) = o_P(1)$ under any $P_{n,1} \in \mathcal{P}_n$. We choose any $P_{n,1} \in \mathcal{P}_n$ and write
\begin{eqnarray}
	\label{decomp2}
	\tilde \theta - \theta_n(P_{n,1}) = \tilde \theta - \theta_n(P_{n,0}) + \theta_n(P_{n,0}) - \theta_n(P_{n,1}).
\end{eqnarray}
The difference on the left hand side and the first difference on the right hand side are $o_P(1)$ under $P_{n,1}$. This implies that $\|\theta_n(P_{n,0}) - \theta_n(P_{n,1})\| = o(1)$. $\blacksquare$ \medskip

\noindent \textbf{Proof of Theorem \ref{thm: cluster dep}: } Let us first show sufficiency. Suppose that either (a) or (b) in Lemma \ref{lemm: equiv} holds. Let us take
\begin{align*}
	\hat \sigma_{LR}^2 = \frac{1}{n}\sum_{m=1}^{M_n} \sum_{i,j \in N_{n,m}} (X_{n,i} - \overline X_n)(X_{n,j} - \overline X_n).
\end{align*}
Since $\sigma_{LR}^2 \le c$ for all $n \ge 1$, we have
\begin{align*}
	\overline X_n = \mathbf{E}[X_{n,i}] +o_P(1).
\end{align*}
Hence
\begin{align*}
	\hat \sigma_{LR}^2 = \frac{1}{n}\sum_{m=1}^{M_n} \sum_{i,j \in N_{n,m}} (X_{n,i} - \mathbf{E}[X_{n,i}])(X_{n,j} - \mathbf{E}[X_{n,i}]) + o_P(1).
\end{align*}
Note that
\begin{align}
	\label{bd}
	\sum_{m=1}^{M_{n}} \left( \frac{n_{m}}{n} \right)^2 \le \sum_{m=1}^{M_{n}} \frac{n_{m}}{n} \le 1.
\end{align}
Choose $P_n \in \mathcal{P}_n$. Then, under $P_n$,
\begin{align}
	\label{dev}
	\mathbf{E}\left[ \left( \hat \sigma_{LR}^2 - \sigma_{LR}^2\right)^2 \right] &= \frac{1}{n^2} \sum_{m=1}^{M_n} \sum_{i,j \in N_{n,m}} \sum_{i',j' \in N_{n,m}} \Cov\left( X_{n,i} X_{n,j}, X_{n,i'} X_{n,j'}\right)\\ \notag
	&\le \frac{C'}{n^2} \sum_{m=1}^{M_n} n_m^2
	= \frac{C'}{n^2} \left((n - n^*) + \sum_{m=1: n_m \ge 2}^{M_n} n_m^2 \right)\\ \notag
	&=  C' \left(\frac{n - n^*}{n^2} + C'\left(\frac{n^*}{n}\right)^2 \sum_{m=1: n_m \ge 2}^{M_n} \left(\frac{n_m}{n^*}\right)^2 \right),
\end{align}
where $C'>0$ is a constant that does not depend on $n$. By (\ref{bd}), the last term is $o(1)$, if either of the conditions (a) and (b) in Lemma \ref{lemm: equiv}. Therefore, $\sigma_{LR}^2$ is consistently estimable in $\mathcal{P}_n$.

Now, let us show necessity. Suppose that both (a) and (b) in Lemma \ref{lemm: equiv} are violated. That is,
\begin{eqnarray}
\label{cond cluster2}
\limsup_{n \rightarrow \infty} \frac{n^*}{n} > 0 \quad\text{and}\quad\limsup_{n \rightarrow \infty} \sum_{m \ge 1: n_m \ge 2}^{M_n} \left(\frac{n_m}{n^*}\right)^2 > 0.
\end{eqnarray}
We fix $\sigma^2$ and show that $\sigma_{LR}^2$ is not consistently estimable in $\mathcal{P}_{n,\mathcal{N}}(\sigma^2)$. We choose $\delta_{n,m}>0$ for each cluster $m$ and $n$ such that (\ref{bound312}) holds for some $\overline\delta \in [-a,a]\setminus \{0\}$, $a< 1/2$, if $n_m \ge 2$. By (\ref{bd}), there exists a subsequence $\{n_k\} \subset \{n\}$ such that
\begin{eqnarray}
\label{conv22}
	\sum_{m=1}^{M_{n_k}} \left( \frac{n_{m,n_k}}{n_k} \right)^2 \rightarrow \tilde c_1 > 0 \quad\text{and}\quad \frac{n_k^*}{n_k} \rightarrow \tilde c_2,
\end{eqnarray}
for some constants $\tilde c_1, \tilde c_2 \in (0,1]$. For simplicity, we fix this subsequence, and denote $n_k$ by $n$.

We let $\Sigma_n$ be the block diagonal $n \times n$ matrix whose $m$-th block is given by $\sigma^2((1-\delta_{n,m})I_{n_m} + \delta_{n,m} \textbf{1}_{n_m} \textbf{1}_{n_m}^\top)$. We show that $\Phi(0,\Sigma_n) \triangleleft \Phi(0,\sigma^2 I_n)$. First, we observe that by Lemma \ref{lemm: unif int}(ii), $\log (d \Phi(0,\Sigma_n)/d \Phi(0,\sigma^2 I_n))(X_n)$ is uniformly tight under $\Phi(0,\sigma^2 I_n)$. Furthermore, we find that by Prohorov's Theorem, there exists a subsequence $\{n_k\}$ of $\{n\}$ such that the sequence $\log (d\Phi(0,\Sigma_{n_k})/d \Phi(0,\sigma^2 I_{n_k}))(X_n)$ weakly converges. Let $W$ be a random variable whose distribution is identical to the weak limit. By the Continuous Mapping Theorem, we have
\begin{eqnarray*}
	\frac{d \Phi(0,\Sigma_{n_k})}{d \Phi(0,\sigma^2 I_{n_k})}(X_{n_k}) \rightarrow_d e^W,
\end{eqnarray*}
along the subsequence $\{n_k\}$. Note that $\mathbf{E}[(d \Phi(0,\Sigma_{n_k})/d\Phi(0,\sigma^2 I_{n_k}))(X_{n_k})] =1 $, where the expectation is under $\Phi(0,\sigma^2 I_{n_k})$. By Lemma \ref{lemm: unif int}(i), $(d \Phi(0,\Sigma_{n_k})/d\Phi(0,\sigma^2 I_{n_k}))(X_{n_k})$ is uniformly integrable under $\Phi(0,\sigma^2 I_{n_k})$. Hence, we find that $\mathbf{E} e^W =1$. By Le Cam's First Lemma \citep[e.g.,][Lemma 6.4]{vanderVaart:98:AsympStat}, we conclude that $\Phi(0,\Sigma_{n_k}) \triangleleft \Phi(0,\sigma^2 I_{n_k})$.

On the other hand, note that the difference between the long-run variances under $\Phi(0,\Sigma_n)$ and under $\Phi(0,\sigma^2 I_n)$ is given by
\begin{align*}
	\frac{\sigma^2 }{n} \sum_{m=1: n_m \ge 2}^{M_n} \mathbf{1}_{n_m}^\top  \Delta_{n,m} \mathbf{1}_{n_m}
	= \sigma^2 \overline \delta \sum_{m=1: n_m \ge 2}^{M_n} \frac{n_m (n_m - 1)}{n n^*} = \sigma^2 \overline \delta \sum_{m=1}^{M_n} \frac{n_m^2}{n n^*}  - \sigma^2 \overline \delta \sum_{m=1: n_m \ge 2}^{M_n} \frac{n_m}{n n^*},
\end{align*}
where $\Delta_{n,m} = \delta_{n,m} \textbf{1}_{n_m} \textbf{1}_{n_m}^\top - \delta_{n,m} I_{n_m}$. We rewrite the last term as
\begin{align*}
	\sigma^2 \overline \delta \sum_{m=1: n_m \ge 2}^{M_n} \frac{n_m^2}{n n^*}  - \frac{\sigma^2 \overline \delta }{n} &= \overline \delta \sum_{m=1: n_m \ge 2}^{M_n} \frac{n_m^2}{n n^*}  +o(1)\\
	&=  \frac{\sigma^2 \overline \delta n^*}{n} \sum_{m=1: n_m \ge 2}^{M_n} \left(\frac{n_m}{n^*}\right)^2 + o(1) \rightarrow \sigma^2 \overline \delta \tilde c_1 \tilde c_2 \ne 0,
\end{align*}
as $n \rightarrow \infty$, where the last convergence is due to (\ref{conv22}). Certainly, $\sigma_{LR}^2$ is consistently estimable along $\Phi(0,\sigma^2 I_n)$. By Lemma \ref{lemm: basic}, we conclude that $\sigma_{LR}^2$ is not consistently estimable in $\mathcal{P}_{n,\mathcal{N}}(\sigma^2)$. Hence, it is not consistently estimable in $\mathcal{P}_{n}$ either. $\blacksquare$\medskip

\noindent \textbf{Proof of Corollary \ref{cor: cluster dep2}: } Let us show sufficiency. First suppose that $\mathcal{M}_n$ consists of negligible clusters, so that $\max_{1 \le m \le M_n} n_m / n \rightarrow 0$, as $n \rightarrow \infty$. From (\ref{bd}), this implies that
\begin{align*}
	\sum_{m=1: n_m \ge 2}^{M_n} \left( \frac{n_m}{n^*}\right)^2 \le \left(\frac{n}{n^*}\right) \max_{1 \le m \le M_n} \frac{n_m}{n} \rightarrow 0,
\end{align*}
as $n \rightarrow \infty$. Therefore, $\sigma_{LR}^2$ is consistently estimable in $\mathcal{P}_{n}$ by Theorem \ref{thm: cluster dep}.

Conversely, suppose that for some $\epsilon>0$,
\begin{align*}
	\limsup_{n \rightarrow \infty} 	\max_{1 \le m \le M_n} \frac{n_m}{n} > \epsilon.
\end{align*}
Then there exist subsequences $\{n_k\} \subset \{n\}$ and $\{n_{m(n_k)}\} \subset \{n_{m(n)}\}_{n \ge 1}$, $m(n) \in \{1,...,M_n\}$, such that
\begin{align*}
	\lim_{k \rightarrow \infty} \frac{n_{m(n_k)}}{n_k} > \epsilon.
\end{align*}
This implies that
\begin{align*}
	\limsup_{n \rightarrow \infty} \sum_{m=1: n_m \ge 2}^{M_n} \left( \frac{n_m}{n^*}\right)^2 > 0.
\end{align*}
Hence, $\sigma_{LR}^2$ is not consistently estimable in $\mathcal{P}_{n}$ by Theorem \ref{thm: cluster dep}. $\blacksquare$\medskip

\noindent \textbf{Proof of Corollary \ref{cor: dep graph}: } (i) Let $N_\circ \subset \{1,...,n\}$ be the set of nodes in the clique with size $n_C$. Take $\mathcal{M}_n$ to be the cluster structure such that there is only one non-singleton cluster that is $N_\circ$. It suffices to show that $\sigma_{LR}^2$ is not consistently estimable in $\mathcal{P}_{n,\mathcal{N}}(\sigma^2)$ with the cluster structure $\mathcal{M}_n$. Note that
\begin{align*}
	\sum_{m=1: n_m \ge 2}^{M_n} \left( \frac{n_m}{n^*}\right)^2 = 1,
\end{align*}
because we have only one non-singleton cluster in $\mathcal{M}_n$. By Theorem \ref{thm: cluster dep}, $\sigma_{LR}^2$ is not consistently estimable in $\mathcal{P}_{n,\mathcal{N}}(\sigma^2)$ with any fixed $\sigma^2>0$. 

(ii) Let us define $\overline N(i) = \{j \in N_n: ij \in \overline E_n\}$, where $\overline E_n = E_n \cup \{ii: i \in N_n\}$, and consider
\begin{eqnarray*}
	\hat \sigma_{LR}^2 = \frac{1}{n}\sum_{i=1}^{n} \sum_{j \in \overline N(i)} (X_{n,i} - \overline X_n) (X_{n,j} - \overline X_n).
\end{eqnarray*}
By rearranging terms, we can write
\begin{align}
	\label{dev2}
	\hat \sigma_{LR}^2 - \sigma_{LR}^2 &= \frac{1}{n}\sum_{i=1}^n \sum_{j: ij \in \overline E_n}(X_{n,i} X_{n,j} - \mathbf{E}[X_{n,i} X_{n,j}])- 2(\overline X_n - \mathbf{E}[X_{n,i}]) \frac{1}{n}\sum_{j \in \overline N(i)} X_{n,j}\\ \notag
	&\quad - 2\mathbf{E}[X_{n,i}] \frac{1}{n}\sum_{i=1}^n \sum_{j \in \overline N(i)} (X_{n,j} - \mathbf{E} X_{n,j}) + \overline X_n^2 - (\mathbf{E}[X_{n,i}])^2 \frac{1}{n}\sum_{i=1}^n |\overline N(i)|.
\end{align}
We can write the squared $L^2$ norm of the leading term on the right hand side as
\begin{align*}
	\frac{1}{n^2}\sum_{i_1=1}^n\sum_{i_2=1}^n \sum_{j_1: i_1j_1 \in \overline E_n} \sum_{j_2: i_2j_2 \in \overline E_n} \Cov\left(X_{n,i_1} X_{n,j_1},X_{n,i_2} X_{n,j_2}\right).
\end{align*}
If $\{i_1,j_1\}$ and $\{i_2,j_2\}$ are not adjacent in $G_n$, the covariance above is zero by the dependency graph assumption. The number of the terms in the above sum such that $\{i_1,j_1\}$ and $\{i_2,j_2\}$ are adjacent in $G_n$ is of the order $O(nd_{mx}^2 d_{av}) = o(n^2)$. The last rate comes from our assumption that $\lim_{n \rightarrow \infty} d_{mx}^2 d_{av} /n = 0$. Therefore, the leading term on the right hand side of (\ref{dev2}) is $o_P(1)$. Similarly, we can show that the remainder terms are $o_P(1)$. Hence, $\hat \sigma_{LR}^2$ is a consistent estimator of $\sigma_{LR}^2$. $\blacksquare$

\bibliographystyle{econometrica}
\bibliography{impossibility}

\end{document}